\documentclass[final,5p,times,twocolumn]{elsarticle}
\usepackage{amssymb}
\usepackage{amsmath}
\usepackage{amsthm}

\newtheorem{definition}{Definition}
\newtheorem{theorem}{Theorem}

\usepackage{algorithm}
\usepackage{algorithmic}

\newcommand{\algofontsize}{\fontsize{7}{8}\selectfont}
\usepackage[hyphens]{url}
\usepackage{hyperref}
\usepackage{breakurl}
\usepackage{makecell}

\usepackage{subfigure}

\usepackage{AMMALanguages}

\usepackage{picins}

\newcommand{\Keywords}{\lstset{keywords={if,then,can,be,active,in,execute}}}

\usepackage{lipsum}
\usepackage{pdfpages}

\newcommand\blfootnote[1]{%
  \begingroup
  \renewcommand\thefootnote{}\footnote{#1}%
  \addtocounter{footnote}{-1}%
  \endgroup
}

\begin{document}

\includepdf[pages={1}]{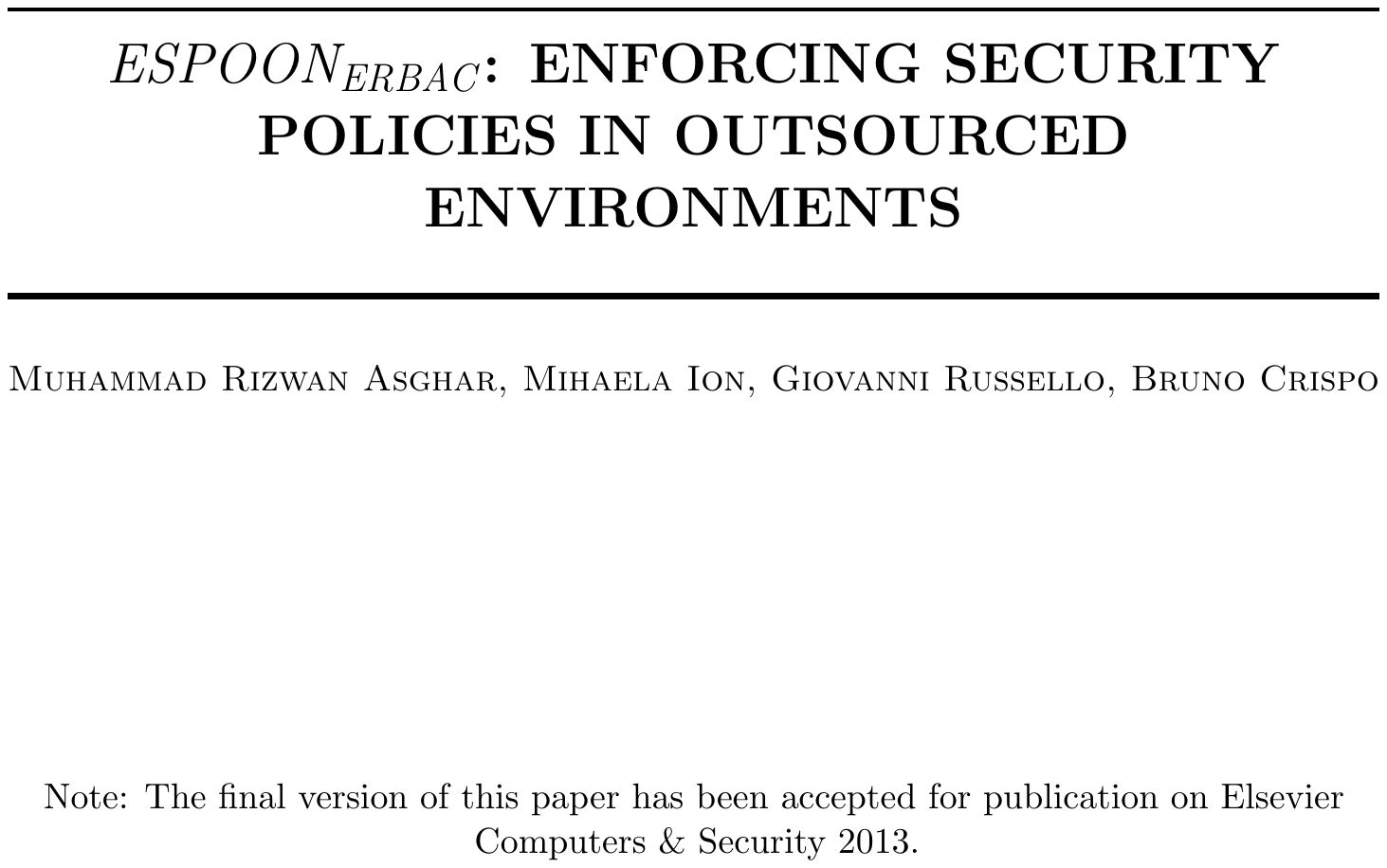}

% left bottom right top
%\includegraphics[trim=50mm 50mm 50mm 0mm,clip,width=\textwidth]{cose-erbac-title}

\mathchardef\mhyphen="2D

\begin{frontmatter}



%\title{$\mathit{ESPOON_{ERBAC}}$: Enforcing Security Policies in \\ Outsourced Environments with Encrypted RBAC}
\title{$\mathit{ESPOON_{ERBAC}}$: Enforcing Security Policies \\ in Outsourced Environments}


\author[cn,disi]{Muhammad Rizwan Asghar}
\ead{asghar@create-net.org}
\author[cn,disi]{Mihaela Ion}
\ead{ion@create-net.org}
\author[nz]{Giovanni Russello}
\ead{g.russello@auckland.ac.nz}
\author[disi]{Bruno Crispo}
\ead{crispo@disi.unitn.it}

\address[cn]{CREATE-NET, International Research Center, Trento Italy}

\address[nz]{Department of Computer Science, The University of Auckland, Auckland New Zealand}

\address[disi]{Department of Information Engineering and Computer Science, University of Trento, Trento Italy}

\begin{abstract}

Data outsourcing is a growing business model offering services to individuals and enterprises for processing and storing a huge amount of data. It is not only economical but also promises higher availability, scalability, and more effective quality of service than in-house solutions. Despite all its benefits, data outsourcing raises serious security concerns for preserving data confidentiality. There are solutions for preserving confidentiality of data while supporting search on the data stored in outsourced environments. However, such solutions do not support access policies to regulate access to a particular subset of the stored data. 

For complex user management, large enterprises employ Role-Based Access Controls (RBAC) models for making access decisions based on the role in which a user is active in. However, RBAC models cannot be deployed in outsourced environments as they rely on trusted infrastructure in order to regulate access to the data. The deployment of RBAC models may reveal private information about sensitive data they aim to protect. In this paper, we aim at filling this gap by proposing \textbf{$\mathit{ESPOON_{ERBAC}}$} for enforcing RBAC policies in outsourced environments. $\mathit{ESPOON_{ERBAC}}$ enforces RBAC policies in an encrypted manner where a curious service provider may learn a very limited information about RBAC policies. We have implemented $\mathit{ESPOON_{ERBAC}}$ and provided its performance evaluation showing a limited overhead, thus confirming viability of our approach.

\end{abstract}


\end{frontmatter}

% \linenumbers

\section{Introduction}
\blfootnote{* The final version of this paper has been accepted for publication in Elsevier Computers \& Security 2013 \cite{Asghar2013-COSE}.}In recent years, data outsourcing has become a very attractive business model. It offers services to individuals and enterprises for processing and storing a huge amount of data at very low cost. It promises higher availability, scalability, and more effective quality of service than in-house solutions. Many sectors including government and healthcare, initially reluctant to data outsourcing, are now adopting it \cite{Ondo2006}.

Despite all its benefits, data outsourcing raises serious security concerns for preserving data confidentiality. The main problem is that the data stored in outsourced environments are within easy reach of service providers that could gain unauthorised access. There are several solutions for guaranteeing confidentiality of data in outsourced environments. For instance, solutions as those proposed in \cite{Dong2011, Kamara2010} offer a protected data storage while supporting basic search capabilities performed on the server without revealing information about the stored data. However, such solutions do not support access policies to regulate the access to a particular subset of the stored data.

\subsection{Motivation}
Solutions for providing access control mechanisms in outsourced environments have mainly focused on encryption techniques that couple access policies with a set of keys, such as the one described in \cite{Vimercati2008}. Only users possessing a key (or a set of hierarchy-derivable keys) are authorised to access the data. The main drawback of these solutions is that security policies are tightly coupled with the security mechanism, thus incurring high processing cost for performing any administrative change for both the users and the policies representing the access rights.

A policy-based solution, such the one described for the Ponder language in \cite{Russello2007}, is more flexible and easy to manage because it clearly separates the security policies from the enforcement mechanism. However, policy-based access control mechanisms are not designed to operate in outsourced environments. Such solutions can work only when they are deployed and operated within a trusted domain (i.e., the computational environment managed by the organisation owning the data). If these mechanisms are outsourced to an untrusted environment, the access policies that are to be enforced on the server may leak information on the data they are protecting. As an example, let us consider a scenario where a hospital has outsourced its healthcare data management services to a third party service provider. We assume that the service provider is honest-but-curious, similar to the existing literature on data outsourcing (such as \cite{Vimercati2007}), i.e., it is honest to perform the required operations as described in the protocol but curious to learn information about stored or exchanged data. In other words, the service provider does not preserve data confidentiality. A patient's medical record should be associated with an access policy in order to prevent an unintended access. The data is stored with an access policy. As an example, let us consider the following access policy: \emph{only a Cardiologist may access the data}. From this policy, it is possible to infer important information about the user's medical conditions (even if the actual medical record is encrypted). This policy reveals that a patient could have heart problems. A misbehaving service provider may sell this information to banks that could deny the patient a loan given her health conditions.

Now-a-days, the most widely used security model is Role-Based Access Controls (RBAC) \cite{Sandhu1996} that makes decision based on role in which a user is active in \cite{Connor2010}. However, the current variants of RBAC model cannot be deployed in outsourced environments as they assume a trusted infrastructure in order to regulate access on data. In RBAC models, RBAC policies may leak information about the data they aim to protect. Asghar \emph{et al.} \cite{Asghar2011ARES} propose $\mathit{ESPOON}$ that aims at enforcing authorisation policies in outsourced environments. They extend $\mathit{ESPOON}$ \cite{Asghar2011ARES} to support RBAC policies and role hierarchies \cite{Asghar2011CCS}. However, they consider that the role assignment is performed by the Company RBAC Manager, which is run in the trusted environment.

\subsection{Research Contributions}
In this paper, we present an RBAC mechanism for outsourced environments where we support full confidentiality of RBAC policies. We named our solution \textbf{$\mathit{ESPOON_{ERBAC}}$} (Enforcing Security Policies in OutsOurced envirOnmeNts with Encrypted RBAC). One of the main advantages of $\mathit{ESPOON_{ERBAC}}$ is that we maintain the clear separation between RBAC policies and the actual enforcing mechanism without loss of policies confidentiality under the assumption that the service provider is honest-but-curious. Our approach allows enterprises to outsource their RBAC mechanisms as a service with all the benefits associated with this business model without compromising the confidentiality of RBAC policies. Summarising, the research contributions of our approach are threefold. First, the service provider does not learn anything about RBAC policies and the requester's attributes during the policy deployment or evaluation processes. Second, $\mathit{ESPOON_{ERBAC}}$ is capable of handling complex contextual conditions (a part of RBAC policies) involving non-monotonic boolean expressions and range queries. Third, the system entities do not share any encryption keys and even if a user is deleted or revoked, the system is still able to perform its operations without requiring re-encryption of RBAC policies. As a proof-of-concept, we have implemented a prototype of our RBAC mechanism and analysed its performance to quantify the overhead incurred by cryptographic operations used in the proposed scheme.

\subsection{Organisation}
The rest of this paper is organised as follows: Section \ref{sec:related-work} reviews the related work. Section \ref{sec:rbac-overview} provides an overview of RBAC models. Section \ref{sec:approach} presents the proposed architecture of $\mathit{ESPOON_{ERBAC}}$. Section \ref{sec:solution-details} and Section \ref{sec:algorithmic-details} focus on solution details and algorithmic details, respectively. Section \ref{sec:security-analysis} provides security analysis of $\mathit{ESPOON_{ERBAC}}$. Section \ref{sec:performance-analysis} analyses the performance overhead of $\mathit{ESPOON_{ERBAC}}$. Finally, Section \ref{sec:conclusions-future-work} concludes this paper and gives directions for the future work.

\iffalse
Section \ref{sec:conclusions-future-work} Conclusions and Future Work
Section \ref{sec:related-work} Related Work
Section \ref{sec:rbac-overview} Overview of RBAC
Section \ref{sec:approach} The $\mathit{ESPOON_{ERBAC}}$ Approach
Section \ref{sec:solution-details} Solution Details
Section \ref{sec:algorithmic-details} Algorithmic Details
Section \ref{sec:security-analysis} Security Analysis
%Section \ref{sec:discussion} Discussion
% Section \ref{sec:discussion} provides the discussion about the security aspects of $\mathit{ESPOON_{ERBAC}}$. 
Section \ref{sec:performance-analysis} Performance Analysis
Section \ref{sec:conclusions-future-work} Conclusions and Future Work
\fi

\section{Related Work}
\label{sec:related-work}

Work on outsourcing data storage to a third party has been focusing on protecting the data confidentiality within the outsourced environment. Several techniques have been proposed allowing authorised users to perform efficient queries on the encrypted data while not revealing information on the data and the query \cite{Song2000, Boneh2004, Golle2004, Curtmola2006, Hwang2007, Boneh2007, Wang2008, Baek2008, Rhee2010, Shao2010, Dong2011}. However, these techniques do not support the case of users having different access rights over the protected data. Their assumption is that once a user is authorised to perform search operations, there are no restrictions on the queries that can be performed and the data that can be accessed.

The idea of using an access control mechanism in an outsourced environment was initially explored in \cite{Vimercati2007-2, Vimercati2007}. In this approach, Vimercati \emph{et al.} provide a selective encryption strategy for enforcing access control policies. The idea is to have a selective encryption technique where each user has a different key capable of decrypting only the resources a user is authorised to access. In their scheme, a public token catalogue expresses key derivation relationships. However, the public catalogue contains tokens in the clear that express the key derivation structure. The tokens could leak information on access control policies and on the protected data. To circumvent the issue of information leakage, in \cite{Vimercati2008} Vimercati \emph{et al.} provide an encryption layer to protect the public token catalogue. This requires each user to obtain the key for accessing a resource by traversing the key derivation structure. The key derivation structure is a graph built (using access key hierarchies \cite{Atallah2009}) from a classical access matrix. There are several issues related to this scheme. First, the algorithm of building key derivation structure is very time consuming. Any administrative actions to update access rights require the users to obtain new access keys derived from the rebuilt key derivation structure and it consequently requires data re-encryption with new access keys. Therefore, the scheme is not very scalable and may be suitable for a static environment where users and resources do not change very often. Second, the scheme does not support complex policies where contextual information may be used for granting access rights. For instance, only specific time and location information associated with an access request may be legitimate to grant access to a user.

Another possible approach for implementing an access control mechanism is protecting the data with an encryption scheme where the keys can be generated from the user's credentials (expressing attributes associated with that user). Although these approaches are not devised particularly for outsourced environments, it is still possible to use them as access control mechanisms in outsourced settings. For instance, a recent work by Narayan \emph{et al.} \cite{Narayan2010} employ the variant of Attribute Based Encryption (ABE) proposed in \cite{Bethencourt2007} (i.e., Ciphertext Policy ABE, or CP-ABE in short) to construct an outsourced healthcare system where patients can securely store their Electronic Health Record (EHR). In their solution, each EHR is associated with a secure search index to provide search capabilities while guaranteeing no information leakage. However, one of the problems associated with CP-ABE is that the access structure, representing the security policy associated with the encrypted data, is not protected. Therefore, a curious storage provider might get information on the data by accessing the attributes expressed in the CP-ABE policies. The problem of having the access structure expressed in cleartext affects in general all the ABE constructions \cite{Sahai2005, Goyal2006, Ostrovsky2007, Bethencourt2007}. Therefore, this mechanism is not suitable for guaranteeing confidentiality of access control policies in outsourced environments.

Asghar \emph{et al.} \cite{Asghar2011ARES} propose $\mathit{ESPOON}$ that aims at enforcing authorisation policies in outsourced environments. In $\mathit{ESPOON}$, a data owner (or someone on the behalf of data owners) may attach an authorisation policy with the data while storing it on the outsourced server. Any authorised requester may get access to the data if she satisfies the authorisation policy associated with that data. However, $\mathit{ESPOON}$ lacks to provide support for RBAC policies. In \cite{Asghar2011CCS}, Asghar \emph{et al.} extended $\mathit{ESPOON}$ to support RBAC policies and role hierarchies. However, in \cite{Asghar2011CCS} the role assignment is performed by the Company RBAC Manager, which is run in the trusted environment. On the other hand, in our current architecture, the role assignment is performed by the service provider running in the outsourced environment. In other words, we have eliminated the need of an additional online-trusted-server i.e., the Company RBAC Manager.

Related to the issue of the confidentiality of the access structure, the hidden credentials scheme presented in \cite{Holt2003} allows one to decrypt ciphertexts while the involved parties never reveal their policies and credentials to each other. Data can be encrypted using an access policy containing monotonic boolean expressions which must be satisfied by the receiver to get access to the data. A passive adversary may deduce the policy structure, i.e., the operators (AND, OR, m-of-n threshold encryption) used in the policy but she does not learn what credentials are required to fulfill the access policy unless she possesses them. Bradshaw \emph{et al.} \cite{Bradshaw2004} extend the original hidden credentials scheme to limit the partial disclosure of the policy structure and speed up the decryption operations. However, in this scheme, it is not easy to support non-monotonic boolean expressions and range queries in the access policy. Last, hidden credentials schemes assume that the involved parties are online all the time to run the protocol.

\section{Overview of RBAC Models}
\label{sec:rbac-overview}
% search: execute permissions
RBAC \cite{Sandhu1996} is an access control model that logically maps well to the job-function specified within an organisation. In the basic RBAC model, a system administrator or a security officer assigns permissions to roles and then roles are assigned to users. A user can make an access request to execute permissions corresponding to a role only if he or she is active in that role. A user can be active in a subset of roles assigned to him/her by making a role activation request. In RBAC, a session keeps mapping of users to roles that are active.

In \cite{Sandhu1996}, Sandhu \emph{et al.} extend the basic RBAC model with role hierarchies for structuring roles within an organisation. The concept of role hierarchy introduces the role inheritance. In the role inheritance, a derived role can inherit all permissions from the base role. The role inheritance incurs extra processing overhead as requested permissions might be assigned to the base role of one in which the user might be active. 

The RBAC model may activate a role or grant permissions while taking into account the context under which the user makes the access request or the role activation request \cite{Kim2007, Joshi2005, Strembeck2004, Neumann2003, Lupu1997}. The RBAC model captures this context by defining contextual conditions. A contextual condition requires certain attributes about the environment or the user making the request. These attributes are contextual information, which may include access time, access date and location of the user who is making the request. The RBAC model grants the request if the contextual information satisfy the contextual conditions.

\begin{figure*} [htp]
\centering
% left bottom right top
%\includegraphics[trim=35mm 45mm 30mm 45mm,clip,width=.9\textwidth]{abstract_picture} %.5
% top left bottom right
\includegraphics[angle=-90,trim=55mm 60mm 55mm 50mm,clip,width=.9\textwidth]{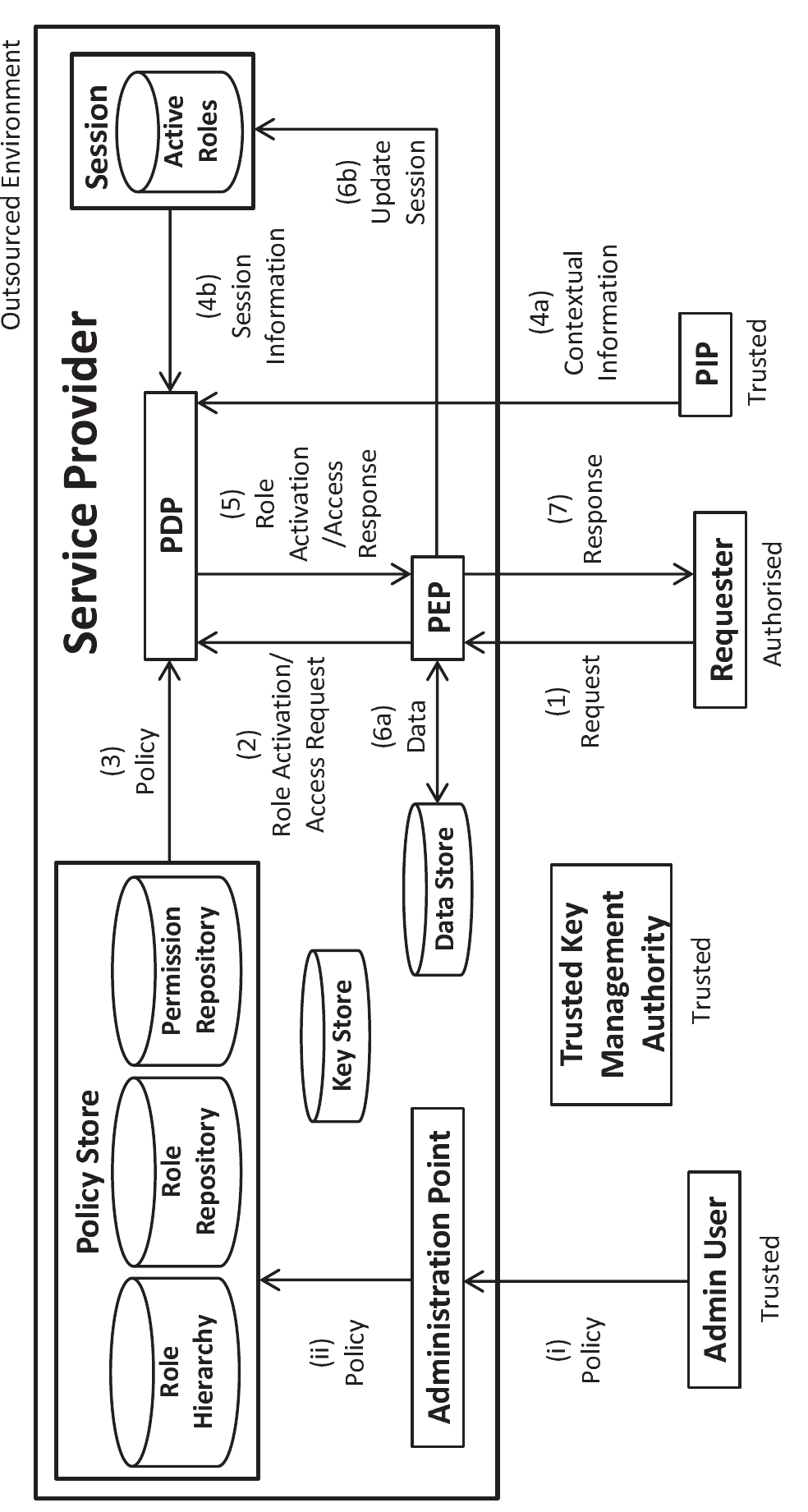}
\caption{The $\mathit{ESPOON_{ERBAC}}$ architecture for enforcing RBAC policies in outsourced environments}
\label{fig:abstract_picture}
\end{figure*}

\section{The $\mathit{ESPOON_{ERBAC}}$ Approach}
\label{sec:approach}

% TODO: adjust \\ on update of this paragraph
$\mathit{ESPOON_{ERBAC}}$ aims at providing RBAC mechanism that can be deployed in an outsourced environment. Figure \ref{fig:abstract_picture} illustrates the proposed architecture that has similar components to the widely accepted architecture for the policy-based management proposed by IETF \cite{Yavatkar2000}. In \\ $\mathit{ESPOON_{ERBAC}}$, an \textbf{Admin User} deploys (i) RBAC policies and sends them to the \textbf{Administration Point} that stores (ii) RBAC policies\footnote{In the rest of this paper, by term \emph{policies} we mean \emph{RBAC policies}.} in the \textbf{Policy Store}. These policies may include permissions assigned to roles, roles assigned to users and the role hierarchy graph that are stored in the Permission Repository, the Role Repository and the Role Hierarchy repository, respectively. 

% role activation
A \textbf{Requester} may send (1) the role activation request to the \textbf{Policy Enforcement Point} (PEP). This request includes the Requester's identifier and the requested role. The PEP forwards (2) the role activation request to the \textbf{Policy Decision Point} (PDP). The PDP retrieves (3) the policy corresponding to the Requester from the Role Repository of the \textbf{Policy Store} and fetches (4) the contextual information from the \textbf{Policy Information Point} (PIP). The contextual information may include the environmental and Requester's attributes under which the requested role can be activated. For instance, consider a contextual condition where a role doctor can only be activated during the duty hours. For simplicity, we assume that the PIP collects all required attributes and sends all of them together in one go. Moreover, we assume that the PIP is deployed in the trusted environment. However, if attributes forgery is an issue, the PIP can request a trusted authority to sign the attributes before sending them to the PDP. The PDP evaluates role assignment policies against the attributes provided by the PIP checking if the contextual information satisfies contextual conditions and sends to the PEP (5) the role activation response. In case of \emph{permit}, the PEP activates the requested role by updating the \textbf{Session} containing the Active Roles repository (6a). Otherwise, in case of \emph{deny}, the requested role is not activated. Optionally, a response can be sent to the Requester (7) with either \emph{success} or \emph{failure}.

% access request
After getting active in a role, a Requester can make the access request that is sent to the PEP (1). This request includes the Requester's identifier, the requested data (target) and the action to be performed. The PEP forwards (2) the access request to the PDP. After receiving the access request, the PDP first retrieves from the Session information about the Requester if she is already active in any role (3a). If so, the PDP evaluates if the Requester's (active) role is permitted to execute the requested action on the requested data. For this purpose, the PDP retrieves (3) the permission assignment policy corresponding to the active role from the Permission Repository of the Policy Store and fetches (4) the contextual information from the PIP required for evaluating contextual conditions in the permission assignment policy. For instance, consider the example where a \emph{Cardiologist} can access the cardiology report during the office hours. The PDP evaluates the permission assignment policies against the attributes provided by the PIP checking if the contextual information satisfies any contextual conditions and sends to the PEP (5) the access response. In case of \emph{permit}, the PEP forwards the access action to the \textbf{Data Store} (6b). In case if no contextual condition is satisfied, the PDP retrieves the role hierarchy from the Role Hierarchy repository of the Policy Store and then traverses this role hierarchy graph in order to find if any base role, the Requester's role might be derived from, has permission to execute the requested action on the requested data. If so, the PEP forwards the access action to the Data Store (6b). Otherwise, in case of \emph{deny}, the requested action is not forwarded. Optionally, a response can be sent to the Requester (7) with either \emph{success} or \emph{failure}.

The main difference with the standard proposed by IETF is that the $\mathit{ESPOON_{ERBAC}}$ architecture is outsourced in an untrusted environment (see Figure \ref{fig:abstract_picture}). The trusted environment comprises only a minimal IT infrastructure that is the applications used by the Admin Users and Requesters, together with the PIP. This reduces the cost of maintaining an IT infrastructure. Having the reference architecture in the cloud increases its availability and provides a better load balancing compared to a centralised approach. In outsourced environments, $\mathit{ESPOON_{ERBAC}}$ guarantees that the confidentiality of policies is protected not only when they are deployed but also when they are enforced. This offers a more efficient evaluation of policies. For instance, a naive solution would see the encrypted policies stored in the cloud and the PDP deployed in the trusted environment. At each evaluation, the encrypted policies would be sent to the PDP that decrypts the policies for a cleartext evaluation. After that, the policies need to be encrypted and send back to the cloud. The \textbf{Service Provider}, where the architecture is outsourced, is honest-but-curious. This means that the provider allows the $\mathit{ESPOON_{ERBAC}}$ components to follow the specified protocols, but it may be curious to find out information about the data and the policies regulating the accesses to the data. As for the data, we assume that data confidentiality is preserved by one of the several techniques available for outsourced environments \cite{Dong2011, Rhee2010, Shao2010}. However, to the best of our knowledge, no solution exists that addresses the problem of guaranteeing the policy confidentiality while allowing an efficient evaluation mechanism that is clearly separated from the policies. Most of the techniques discussed in the related work section require the security mechanism to be tightly coupled with the policies. In the following section, we can show that it is possible to maintain a generic PDP separated from the security policies and able to take access decisions based on the evaluation of encrypted policies. In this way, the policy confidentiality can be guaranteed against a curious provider and the functionality of the access control mechanism is not restricted.

\subsection{System Model}
Before presenting the detail of the scheme used in $\mathit{ESPOON_{ERBAC}}$, it is necessary to discuss the system model. In this section, we identify the following system entities:

\begin{itemize}

\item \textbf{Admin User:} This type of user is responsible for the administration of policies stored in the outsourced environment. An Admin User can deploy new policies or update/delete already deployed policies.

\item \textbf{Requester:} A Requester is a user that requests an access (e.g., read, write or search) over the data residing in the outsourced environment. Before the access is permitted, policies deployed in the outsourced environment are evaluated.

\item \textbf{Service Provider (SP):} The SP is responsible for managing the outsourced computation environment, where the $\mathit{ESPOON_{ERBAC}}$ components are deployed and to store the data, and policies. It is assumed the SP is honest-but-curious (as \cite{Vimercati2007} does), i.e., it allows the components to follow the protocol to perform the required actions but curious to deduce information about the exchanged and stored policies.

\item \textbf{Trusted Key Management Authority (TKMA):} The TKMA is fully trusted and responsible for generating and revoking the keys. For each type of authorised users (including an Admin User and a Requester), the TKMA generates two key sets and securely transmits the client key set to the user and the server key set to the Administration Point. The Administration Point inserts the server side key set in the \textbf{Key Store}. The TKMA is deployed on the trusted environment. Although requiring a TKMA seems at odds with the need of outsourcing the IT infrastructure, we argue that the TKMA requires less resources and less management effort. Securing the TKMA is much easier since a very limited amount of data needs to be protected and the TKMA can be kept offline most of the time.

\end{itemize}

It should be clarified that in our settings an Admin User is not interested in protecting the confidentiality of policies from other Admin Users and Requesters. Here, the main goal is to preserve the confidentiality of data and policies from the SP.

\begin{figure}
\Keywords
% \begin{lstlisting}[style=AMMA,breaklines,mathescape,rulesepcolor=\color{black}]
\begin{lstlisting}[style=AMMA,numbers=none,breaklines,mathescape,rulesepcolor=\color{black}]
if $\langle \mathit{CONDITION} \rangle$ then $\langle USER \rangle$ can be active in $\langle \{ R_1, R_2, \ldots, R_n \} \rangle$

\end{lstlisting}
\caption{RBAC Policy: Role assignment}
\label{fig:policy-role-assignment}
\end{figure}

\begin{figure}
\Keywords
\begin{lstlisting}[style=AMMA,numbers=none,breaklines,mathescape,rulesepcolor=\color{black}]
if $\langle \mathit{CONDITION} \rangle$ then $\langle R \rangle$ can execute $\langle \{ (A_1, T_1), (A_2, T_2), \ldots, (A_n, T_n) \} \rangle$

\end{lstlisting}
\caption{RBAC Policy: Permission assignment}
\label{fig:policy-permission-assignment}
\end{figure}

\subsection{Representation of RBAC Policies/Requests}
\label{sec:representation}
In this section, we provide details about how to represent policies and requests used in our approach. An RBAC policy contains a role assignment policy, a permission policy and a role hierarchy graph. In the following, we discuss each of them. Figure \ref{fig:policy-role-assignment} illustrates how we represent role assignment policies in $\mathit{ESPOON_{ERBAC}}$. The meaning of role assignment policy is as follows: if contextual condition, $\mathit{CONDITION}$, is $\mathit{true}$ then $USER$ can be active in any role(s) out of role set $\{ R_1, R_2, \ldots, R_n \}$. Figure \ref{fig:policy-permission-assignment} illustrates how we represent permission assignment policies in $\mathit{ESPOON_{ERBAC}}$. The meaning of permission assignment policy is as follows: if contextual condition, $\mathit{CONDITION}$, is $\mathit{true}$ then role $R$ can execute any permission(s) out of permission set $\{ (A_1, T_1), (A_2, T_2), \ldots, (A_n, T_n) \}$.

\begin{figure}
\centering
% left bottom right top
\includegraphics[trim=70mm 60mm 35mm 135mm,clip,width=.4\textwidth]{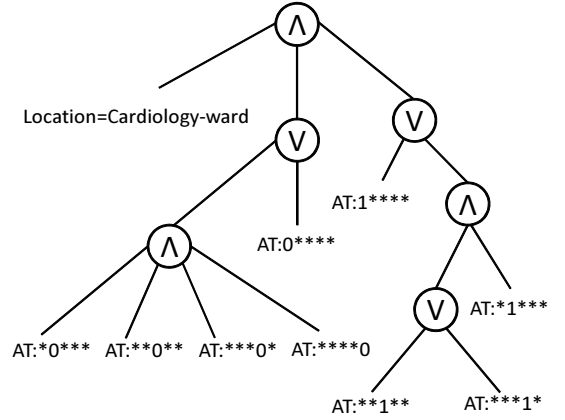}
\caption{An example of contextual condition illustrating $Location=Cardiology \mhyphen ward$ and $AT>9\#5$ and $AT<17\#5$}
\label{fig:cc}
\end{figure}

% CONDITION
% TODO: define CP-ABE if not done so
The PDP evaluates contextual conditions of both role assignment and permission assignment policies before granting the access. In order to evaluate a contextual condition, the PDP requires contextual information. The contextual information captures the context in which a Requester makes access or role activation requests. The PIP collects and sends required contextual information to the PDP. To represent contextual conditions, we use the tree structure described in \cite{Bethencourt2007} for CP-ABE policies. This tree structure allows an Admin User to express contextual conditions as conjunctions and disjunctions of equalities and inequalities. Internal nodes of the tree structure are AND, OR or threshold gates (e.g., 2 of 3) and leaf nodes are values of condition predicates either string or numerical. In the tree structure, a string comparison is represented by a single leaf node. However, the tree structure uses the \emph{bag of bits} representation to support comparisons between numerical values that could express time, date, location, age, or any numerical identifier. For instance, let us consider a contextual condition stating that the Requester location should be $Cardiology \mhyphen ward$ and that the access time should be between 9:00 and 17:00 hrs. Figure \ref{fig:cc} illustrates the tree structure representing this contextual condition, where access time (AT) is in a 5-bit representation (\#5).

% requests
A Requester can make a role activation request $\mathit{ACT}$ or an access request $\mathit{REQ}$. In $ACT = (i, R)$, a Requester includes her identity $i$ along with role $R$ to be activated. After a Requester is active in $R$, she can execute permissions assigned to $R$. For executing any permission, a Requester sends $REQ = (R, A, T)$ that includes $R$ she is active in, action $A$ to be taken over target $T$. A Requester sends $\mathit{ACT}$ or $\mathit{REQ}$ requests to the PEP. 

% PIP
% TODO: adjust \\ on update of this paragraph
The PEP receives and forwards requests $\mathit{ACT}$ or $\mathit{REQ}$ to the PDP. The PDP fetches policies corresponding to requests from the Policy Store. The PDP may require contextual information in order to evaluate contextual conditions to grant $\mathit{ACT}$ or $\mathit{REQ}$. Let us consider $\mathit{CONDITION}$ illustrated in Figure \ref{fig:cc} requiring location of Requester and access time. We assume the Requester makes the request when she is in $\mathit{Cardiology \mhyphen ward}$ and access time (AT) is 10:00 hrs. The PIP collects and then transforms this contextual information as follows: $\mathit{Location = Cardiology \mhyphen ward}$, $\mathit{AT : 0****}$, $\mathit{AT: *1***}$, $\mathit{AT: **0**}$, $\mathit{AT: ***1*}$, \\ $\mathit{AT: ****0}$, where AT is in a 5-bit representation (same as it is in $\mathit{CONDITION}$). After performing transformation, the PIP sends contextual information to the PDP. The PDP receives contextual information and then evaluates $\mathit{CONDITION}$ by first matching attributes in contextual information against leaf-nodes in the $\mathit{CONDITION}$ tree and then evaluating internal nodes according to AND and OR gates.

\begin{figure}
\Keywords
\begin{lstlisting}[style=AMMA,numbers=none,breaklines,mathescape,rulesepcolor=\color{black}]
$R_1$ extends $\langle \{ R_{i}, R_{ii}, \ldots, R_{k_1} \} \rangle$
$R_2$ extends $\langle \{ R_{i}, R_{ii}, \ldots, R_{k_2} \} \rangle$
$\vdots$
$R_n$ extends $\langle \{ R_{i}, R_{ii}, \ldots, R_{k_n} \} \rangle$

\end{lstlisting}
\caption{RBAC Policy: Role hierarchy}
\label{fig:policy-role-hierarchy}
\end{figure}

\begin{figure}
\centering
% left bottom right top
\includegraphics[trim=60mm 35mm 30mm 125mm,clip,width=.3\textwidth]{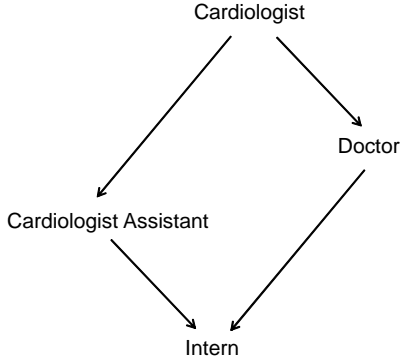} % .2
\caption{Role hierarchy graph}
\label{fig:role_hierarchy_graph}
\end{figure}

% role hierarchy graph
The $\mathit{ESPOON_{ERBAC}}$ architecture supports role inheritance. In role inheritance, a derived role can execute all permissions from its base role. Before denying $\mathit{REQ}$, the PDP may need to check if base role of one in $\mathit{REQ}$ can execute requested permissions. In order to find base roles, we store a role hierarchy graph on the SP. In $\mathit{ESPOON_{ERBAC}}$, the PDP traverses in the role hierarchy graph to find base roles. Figure \ref{fig:policy-role-hierarchy} illustrates how we represent a role hierarchy graph. In Figure \ref{fig:policy-role-hierarchy}, each line represents a role that may extend a set of roles. All these inheritance rules may form a role hierarchy graph. For instance, consider an example from healthcare domain where a \emph{Cardiologist Assistant} extends \emph{Intern}, a \emph{Doctor} extends \emph{Intern} and finally a \emph{Cardiologist} extends both \emph{Cardiologist Assistant} and \emph{Doctor}. If we combine all these inheritance rules then it can form a graph as shown in Figure \ref{fig:role_hierarchy_graph}.

% why we need encryption of policies/requests
In this representation, leaf-nodes in $\mathit{CONDITION}$, $R$, $A$, $T$ of both $\mathit{ACT}$ and $\mathit{REQ}$, roles in the role hierarchy graph, and attributes in contextual information are in cleartext. Therefore, such information is easily accessible in the outsourced environment and may leak information about the data that policies protect. In the following, we show how we protect such representation while allowing the PDP to evaluate policies against requests and contextual information.

\section{Solution Details}
\label{sec:solution-details}

$\mathit{ESPOON_{ERBAC}}$ aims at enforcing policies in outsourced environments. The main idea of our approach is to use an encryption scheme for preserving confidentiality of policies while allowing the PDP to perform the correct evaluation. In $\mathit{ESPOON_{ERBAC}}$, we can notice that the operation performed by the PDP for evaluating policies (against attributes in the request and contextual information) is similar to the search operation executed in a database. In particular, in our case the policy is a query; while, attributes in the request ($\mathit{ACT}$ or $\mathit{REQ}$) and contextual information represent the data.

For $\mathit{ESPOON_{ERBAC}}$, as a starting point we consider the multiuser Searchable Data Encryption (SDE) scheme proposed by Dong \emph{et al.} in \cite{Dong2011}. The SDE scheme allows an untrusted server to perform searches over encrypted data without revealing to the server information on both the data and elements used in the request. The advantage of this method is that it offers multi-user access without requiring key sharing between users. Each user in the system has a unique set of keys. The data encrypted by one user can be decrypted by any other authorised user. However, the SDE implementation in \cite{Dong2011} is only able to perform keyword comparison based on equalities. One of the major extensions of our implementation is that we are able to support the evaluation of contextual conditions containing complex boolean expressions such as non-conjunctive and range queries in multi-user settings.

In general, we distinguish four phases in $\mathit{ESPOON_{ERBAC}}$ for managing life cycle of policies in outsourced environments. These phases include \emph{initialisation}, \emph{\textbf{policy deployment}}, \emph{\textbf{policy evaluation}} and \emph{user revocation}. In the following, we provide details of each phase.

\subsection{Initialisation Phase}

In $\mathit{ESPOON_{ERBAC}}$, each user (including an Admin User and a Requester) obtains a client side key from the TKMA while the SP (as a proxy server) receives a server side key set corresponding to the user. The client side key set serves as a private key for a user. The SP stores all key sets in the Key Store. The Key Store is accessible to the Administration Point, the PEP and the PDP.

\subsection{Policy Deployment Phase}

For deploying (or updating existing) policies, an Admin User performs a first round of encryption using her client side key set. An Admin User encrypts elements of policies. In role assignment policies, an Admin User encrypts all roles assigned to a user. In permission assignment policies, an Admin User encrypts both action and target parts of each permission and also encrypts the role to which these permissions are assigned. As we know that a tree represents condition conditions of both role assignment and permission assignment policies (as shown in Figure \ref{fig:cc}), an Admin User encrypts each leaf node of the tree while non-leaf (internal) nodes representing AND, OR or threshold gates are in cleartext. In a role hierarchy graph (as shown in Figure \ref{fig:role_hierarchy_graph}), an Admin User encrypts each of its node representing a role. After completing the first round of encryption on policies, an Admin User sends client encrypted policies to the Administration Point on the SP. These client encrypted policies are protected but cannot be enforced as these are not in common format. To convert client encrypted policies to common format, the Administration Point performs a second round of encryption using server side key set corresponding to the Admin User. The second round of encryption serves as a proxy re-encryption. In the second round of encryption, the Administration Point encrypts all elements that are encrypted in the first round of encryption. Finally, the Administration Point stores server encrypted policies in the Policy Store.

\subsection{Policy Evaluation Phase}

A Requester can make a role activation request $\mathit{ACT}$. Before sending $\mathit{ACT}$ to the SP, a Requester generates a client trapdoor of the role in $\mathit{ACT}$. A Requester generates client trapdoor using her client side key set. The trapdoor representation does not leak information on elements of requests. Similarly, a Requester can make an access request $\mathit{REQ}$ after getting active in a role. A Requester generates a client trapdoor for each element in $\mathit{REQ}$ including the role, the action and the target. A Requester sends requests containing client generated trapdoors to the PEP on the SP. The PEP performs another round of trapdoor generation for converting all trapdoors into a common format. After performing a second round of trapdoor generation on the server side, the PEP forwards server generated trapdoors to the PDP. The PDP fetches policies from the Policy Store and then performs encrypted matching of trapdoors in request against encrypted elements in policies. The encrypted matching in outsourced environments does not leak information about elements of requests or policies. 

The PDP may require contextual information in order to evaluate the contextual conditions of policies. The PIP collects contextual information and generates client trapdoors for elements of contextual information using her client side key set. The PIP sends client generated trapdoors of contextual information to the PDP. The PDP performs another round of trapdoor generation using server side key set corresponding to the PIP. Finally, the PDP evaluates the contextual condition by matching trapdoors of contextual information against encrypted leaf nodes of the tree representing the contextual condition (as shown in Figure \ref{fig:cc}). After evaluating leaf nodes, the PDP evaluates non-leaf nodes of the tree based on AND, OR and threshold gates. The PDP grants the access request if (the root node of) the tree evaluates to $\mathit{true}$.

The PDP may need to find base roles corresponding to the role in $\mathit{REQ}$ considering the fact that a derived role has all permissions from its base role. In order to find base role, the PDP fetches the role hierarchy graph from the Policy Store. The PDP matches trapdoor of role in $\mathit{REQ}$ against server encrypted roles in the role hierarchy graph. While deploying the role hierarchy graph, we store also server generated trapdoor of the role along with each server encrypted of role because the PDP needs a trapdoor of each base role so that it can match this trapdoor against roles in the Permission Repository. After traversing in the role hierarchy graph, the PDP extracts server generated trapdoors of all base roles of one that matches with trapdoor of role in $\mathit{REQ}$. The PDP verifies if any base role has requested permissions. If so, the PDP grants the request.

\subsection{User Revocation Phase}

In $\mathit{ESPOON_{ERBAC}}$, users do not share any keys and a compromised user can be revoked without requiring re-encryption of policies or re-distribution of keys. For revoking a compromised user, the Administration Point removes the server side key set (corresponding to the user) from the Key Store.

% TODO: switch to discussion section?
%deactivate role by sending or a server may deactivate after a certain time.

% system init

\begin{algorithm}[htp]
{\algofontsize
\caption{\textbf{Init}}

\label{algo:init}

\begin{algorithmic}[1]

\REQUIRE A security parameter $1^k$.

\ENSURE The public parameters $param$ and the master secret key $msk$.

\medskip

\STATE Generate primes p and q of size $1^k$ such that $q$ $|$ $p - 1$ \label{line:primes}
\STATE Create a generator $g$ such that $\mathbb{G}$ is the unique order $q$ subgroup of $\mathbb{Z}^*_p$ \label{line:generator}
\STATE Choose a random $x \in \mathbb{Z}^*_q$ \label{line:master-x}
\STATE $h \leftarrow g^x$ \label{line:params-h}
\STATE Choose a collision-resistant hash function $H$ \label{line:params-H}
\STATE Choose a pseudorandom function $f$ \label{line:params-f}
\STATE Choose a random key $s$ for $f$ \label{line:master-s}
\STATE $param \leftarrow (\mathbb{G}, g, q, h, H, f)$ \label{line:params}
\STATE $msk \leftarrow (x, s)$ \label{line:master}

\RETURN $(param, msk)$

\end{algorithmic}
}
\end{algorithm}

\section{Algorithmic Details}
\label{sec:algorithmic-details}

In this section, we provide details of algorithms used in each phase for managing life cycle of policies. All these algorithms constitute the proposed schema. 

\subsection{Initialisation Phase}

In this phase, the system is initialised and then the TKMA generates required keying material for entities in $\mathit{ESPOON_{ERBAC}}$. During the system initlisation, the TKMA takes a security parameter $k$ and outputs the public parameters $params$ and the master key set $msk$ by running \textbf{Init} illustrated in Algorithm \ref{algo:init}. The detail of \textbf{Init} is as follows: the TKMA generates two prime numbers $p$ and $q$ of size $k$ such that $q$ divides $p-1$ (Line \ref{line:primes}). Then, it creates a cyclic group $\mathbb{G}$ with a generator $g$ such that $\mathbb{G}$ is the unique order $q$ subgroup of $\mathbb{Z}^*_p$ (Line \ref{line:generator}). Next, it randomly chooses $x \in \mathbb{Z}^*_q$ (Line \ref{line:master-x}) and compute $h$ as $g^x$ (Line \ref{line:params-h}). Next, it chooses a collision-resistant hash function $H$ (Line \ref{line:params-H}), a pseudorandom function $f$ (Line \ref{line:params-f}) and a random key $s$ for $f$ (Line \ref{line:master-s}). Finally, it publicises the public parameters $params = (\mathbb{G}, g, q, h, H, f)$ (Line \ref{line:params}) and keeps securely the master secret key $msk = (x, s)$ (Line \ref{line:master}).

% key gen

\begin{algorithm}[htp]
{\algofontsize
\caption{\textbf{KeyGen}}

\label{algo:keygen}

\begin{algorithmic}[1]

\REQUIRE The master secret key $msk$, the user identity $i$ and the public parameters $params$.

\ENSURE The client side key set $K_{u_i}$ and server side key set $K_{s_i}$.

\medskip

\STATE Choose a random $x_{i1} \in \mathbb{Z}^*_q$ \label{line:xi1}
\STATE $x_{i2} \leftarrow x - x_{i1}$ \label{line:xi2}
\STATE $K_{u_i} \leftarrow (x_{i1}, s)$ \label{line:ku}
\STATE $K_{s_i} \leftarrow (i, x_{i2})$ \label{line:ks}

\RETURN $(K_{u_i}, K_{s_i})$

\end{algorithmic}
}
\end{algorithm}

For each user (including an Admin User and a Requester), the TKMA generates the keying material. For generating the keying material, the TKMA takes the master secret key $msk$, the user identity $i$ and the public parameters $params$ and outputs two key sets: the client side key set $K_{u_i}$ and the server side key set $K_{s_i}$ by running \textbf{KeyGen} illustrated in Algorithm \ref{algo:keygen}. In \textbf{KeyGen}, TKMA randomly chooses $x_{i1} \in \mathbb{Z}^*_q$ (Line \ref{line:xi1}) and computes $x_{i2} = x - x_{i1}$ (Line \ref{line:xi2}). It creates the client side key set $K_{u_i} = (x_{i1}, s)$ (Line \ref{line:ku}) and the server side key set $K_{s_i} = (i, x_{i2})$ (Line \ref{line:ks}).

% graphical representation of key gen

\begin{figure}
\centering
% left bottom right top
\includegraphics[trim=75mm 60mm 45mm 135mm,clip,width=.4\textwidth]{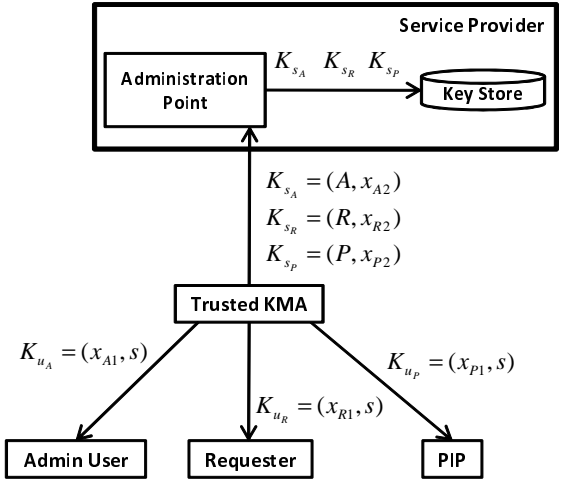} % .3
\caption{Key distribution}
\label{fig:key_generation}
\end{figure}

After running Algorithm \ref{algo:keygen}, the TKMA sends the client side key set $K_{u_i}$ and the server side key set $K_{s_i}$ to user $i$ and the Administration Point on the SP, respectively. The client side key set $K_{u_i}$ serves as a private key for user $i$. The Administration Point of the SP inserts $K_{s_i}$ in the Key Store by updating it as follows: $KS = KS \cup K_{s_i}$. The Key Store is initialised as: $KS \leftarrow \phi$. Figure \ref{fig:key_generation} illustrates key distribution where Admin User $A$, Requester $R$ and PIP $P$ receive $K_{u_A}$, $K_{u_R}$ and $K_{u_P}$, respectively. The TKMA sends the corresponding server side key sets $K_{s_A}$, $K_{s_R}$ and $K_{s_P}$ to the Administration Point on the SP. The Administration Point inserts server side key sets into the Key Store. Please note that only the Administration Point, the PDP and the PEP are authorised to access the Key Store.

% aux method: client enc

\begin{algorithm}[htp]
{\algofontsize
\caption{\textbf{ClientEnc}}

\label{algo:client-enc}

\begin{algorithmic}[1]

\REQUIRE Element $e$, the client side key set $K_{u_i}$ corresponding to Admin User $i$ and the public parameters $params$.

\ENSURE The client encrypted element $c^*_i (e)$.

\medskip

\STATE Choose a random $r_e \in \mathbb{Z}^*_q$ \label{line:ce-r}
\STATE ${\sigma}_e \leftarrow f_s (e)$ \label{line:ce-sigma}
\STATE $\hat{c}_1 \leftarrow g^{r_e+{\sigma}_e}$ \label{line:ce-c1}
\STATE $\hat{c}_2 \leftarrow \hat{c}_1^{x_{i1}}$ \label{line:ce-c2}
\STATE $\hat{c}_3 \leftarrow H(h^{r_e})$ \label{line:ce-c3}
\STATE $c^*_i (e) \leftarrow (\hat{c}_1, \hat{c}_2, \hat{c}_3)$ \label{line:ce-c}

\RETURN $c^*_i (e)$

\end{algorithmic}
}
\end{algorithm}

% aux method: server re enc

\begin{algorithm}[htp]
{\algofontsize
\caption{\textbf{ServerReEnc}}

\label{algo:server-re-enc}

\begin{algorithmic}[1]

\REQUIRE The client encrypted element $c^*_i (e)$ and the server side key set $K_{s_i}$ corresponding to Admin User $i$.

\ENSURE The server encrypted element $c(e)$.

\medskip

\STATE $c_1 \leftarrow (\hat{c}_1)^{x_{i2}}.\hat{c}_2 = \hat{c}_1^{x_{i1}+x_{i2}} = (g^{r_e+{\sigma}_e})^x = h^{r_e+{\sigma}_e}$ \label{line:se-c1}
\STATE $c_2 = \hat{c}_3 = H(h^{r_e})$ \label{line:se-c2}
\STATE $c(e) = (c_1, c_2)$ \label{line:se-c}

\RETURN $c(e)$

\end{algorithmic}
}
\end{algorithm}

% policy deployment 

\begin{figure}
\centering
% left bottom right top
\includegraphics[trim=70mm 60mm 45mm 130mm,clip,width=.4\textwidth]{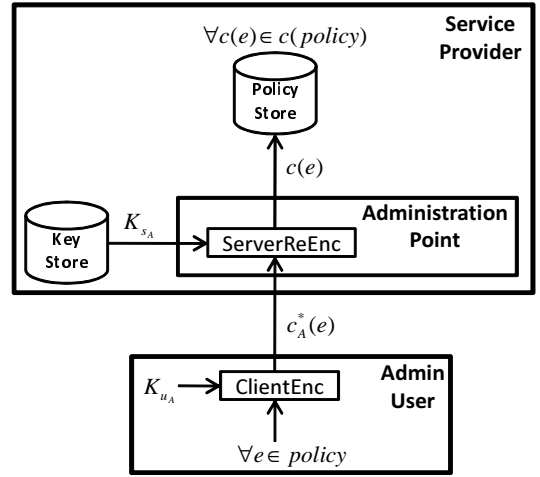} % .3
\caption{Policy deployment phase}
\label{fig:patient}
\end{figure}

\subsection{Policy Deployment Phase}
In the policy deployment phase, an Admin User defines and deploys policies. In general, a policy can be deployed after performing two rounds of encryptions. An Admin User performs a first round of encryption while the Administration Point on the SP performs a second round of encryption. For performing a first round of encryption, an Admin User runs \textbf{ClientEnc} illustrated in Algorithm \ref{algo:client-enc}. \textbf{ClientEnc} takes as input (policy) element $e$, the client side key set $K_{u_i}$ corresponding to Admin User $i$ and the public parameters $params$ and outputs the client encrypted element $c^*_i (e)$. In \textbf{ClientEnc}, an Admin User randomly chooses $r_{e} \in \mathbb{Z}^*_q$ (Line \ref{line:ce-r}), computes ${\sigma}_e$ as $f_s (e)$ (Line \ref{line:ce-sigma}), and then computes $\hat{c}_1$, $\hat{c}_2$ and $\hat{c}_3$ as $g^{r_e+{\sigma}_e}$ (Line \ref{line:ce-c1}), $\hat{c}_1^{x_{i1}}$ (Line \ref{line:ce-c2}) and $H(h^{r_e})$ (Line \ref{line:ce-c3}), respectively. $\hat{c}_1$, $\hat{c}_2$ and $\hat{c}_3$ constitute $c^*_i (e)$ (Line \ref{line:ce-c}). An Admin User transmits to the Administration Point the client encrypted elements of a policy as shown in Figure \ref{fig:patient}.

The Administration Point retrieves the server side key set corresponding to the Admin User and performs a second round of encryption by running \textbf{ServerReEnc} illustrated in Algorithm \ref{algo:server-re-enc}. \textbf{ServerReEnc} takes as input the client encrypted element $c^*_i (e)$ and the server side key set $K_{s_i}$ corresponding to Admin User $i$ and outputs the server encrypted element $c(e)$. The Administration Point calculates $c_1$ and $c_2$ as $(\hat{c}_1)^{x_{i2}}.\hat{c}_2 = \hat{c}_1^{x_{i1}+x_{i2}} = (g^{r_e+{\sigma}_e})^x = h^{r_e+{\sigma}_e}$ (Line \ref{line:se-c1}) and $\hat{c}_3 = H(h^{r_e})$ (Line \ref{line:se-c2}), respectively. Both $c_1$ and $c_2$ form $c(e)$ (Line \ref{line:se-c}). The Administration Point stores the server encrypted policies in the Policy Store as shown in Figure \ref{fig:patient}.

In the following, we describe how to deploy different (parts of) policies including role assignment, permission assignment, contextual conditions and role hierarchy graph. For the deployment of each (part of) policy, we follow general strategy as already described in this section and also illustrated in Figure \ref{fig:patient}.

% deploy: role assignment: client side

\begin{algorithm}[htp]
{\algofontsize
\caption{\textbf{RoleAssignment:ClientSide}}

\label{algo:deploy-role-assignment-client-side}

\begin{algorithmic}[1]

\REQUIRE List of roles $L$ to be assigned to Requester $j$, the client side key set $K_{u_i}$ corresponding to Admin User $i$ and the public parameters $params$.

\ENSURE The client encrypted role assignment list $L_{C_i}$.

\medskip

\STATE $L_{C_i} \leftarrow \phi$ \label{line:deploy-ra-cs-init}

\FOR {each role $r$ in list $L$} \label{line:deploy-ra-cs-loop}

	\STATE $c^*_i (r) \leftarrow$ call \textbf{ClientEnc} ($r$, $K_{u_i}$, $params$) {\algofontsize \COMMENT{see Algorithm \ref{algo:client-enc}}} \label{line:deploy-ra-cs-call-enc}
	\STATE $L_{C_i} \leftarrow L_{C_i} \cup c^*_i (r)$ \label{line:deploy-ra-cs-update}

\ENDFOR

\RETURN ($j$, $L_{C_i}$)

\end{algorithmic}
}
\end{algorithm}

% deploy: role assignment: server side

\begin{algorithm}[htp]
{\algofontsize
\caption{\textbf{RoleAssignment:ServerSide}}

\label{algo:deploy-role-assignment-server-side}

\begin{algorithmic}[1]

\REQUIRE The client encrypted role assignment list $L_{C_i}$ for Requester $j$ and identity $i$ of Admin User.

\ENSURE The server encrypted role assignment list $L_{S}$.

\medskip

\STATE $K_{s_i} \leftarrow KS[i]$ {\algofontsize \COMMENT{retrieve the server side key corresponding to Admin User $i$}} \label{line:deploy-ra-ss-ks}

\STATE $L_{S} \leftarrow \phi$ \label{line:deploy-ra-ss-init}

\FOR {each client encrypted role $c^*_i (r)$ in list $L_{C_i}$} \label{line:deploy-ra-ss-loop}

	\STATE $c(r) \leftarrow$ call \textbf{ServerReEnc} ($c^*_i (r)$, $K_{s_i}$) {\algofontsize \COMMENT{see Algorithm \ref{algo:server-re-enc}}} \label{line:deploy-ra-ss-call}
	
	\STATE $L_{S} \leftarrow L_{S} \cup c(r)$ \label{line:deploy-ra-ss-update}

\ENDFOR

\RETURN ($j$, $L_{S}$)

\end{algorithmic}
}
\end{algorithm}

\emph{\textbf{Deployment of Role Assignment Policies:}} 
In order to assign roles to a Requester, an Admin User can deploy role assignment policies. For this purpose, an Admin User runs \textbf{RoleAssignment:ClientSide} illustrated in Algorithm \ref{algo:deploy-role-assignment-client-side}. This algorithm takes as input a list of roles $L$ to be assigned to Requester $j$, the client side key set $K_{u_i}$ corresponding to Admin User $i$ and the public parameters $params$ and outputs the client encrypted role assignment list $L_{C_i}$. First, it creates and then initialises new list $L_{C_i}$ (Line \ref{line:deploy-ra-cs-init}). For each role in $L$ (Line \ref{line:deploy-ra-cs-loop}), it generates client encrypted role by calling \textbf{ClientEnc} illustrated in Algorithm \ref{algo:client-enc} (Line \ref{line:deploy-ra-cs-call-enc}) and then it updates $L_{C_i}$ by adding client encrypted role (Line \ref{line:deploy-ra-cs-update}). An Admin User sends the client encrypted role assignment list to the Administration Point. 
During the second round of encryption, the Administration Point runs \textbf{RoleAssignment:ServerSide} illustrated in Algorithm \ref{algo:deploy-role-assignment-server-side}. This algorithm takes as input the client encrypted role assignment list $L_{C_i}$ for Requester $j$ and identity $i$ of Admin User and ouputs the server encrypted role assignment list $L_{S}$. While running \textbf{RoleAssignment:ServerSide}, the Administration Point first retrieves the server side key $K_{s_i}$ corresponding to Admin User $i$ (Line \ref{line:deploy-ra-ss-ks}). It creates and initialises new list $L_{S}$ (Line \ref{line:deploy-ra-ss-init}). For each role in $L_{C_i}$ (Line \ref{line:deploy-ra-ss-loop}), it generates server encrypted role by calling \textbf{ServerReEnc} illustrated in Algorithm \ref{algo:server-re-enc} (Line \ref{line:deploy-ra-ss-call}) and updates $L_{S}$ by adding the server encrypted role (Line \ref{line:deploy-ra-ss-update}).

% deploy: permission assignment: client side

\begin{algorithm}[htp]
{\algofontsize
\caption{\textbf{PermissionAssignment:ClientSide}}

\label{algo:deploy-permission-assignment-client-side}

\begin{algorithmic}[1]

\REQUIRE List of permissions $L$ to be assigned to role $r$, the client side key set $K_{u_i}$ corresponding to Admin User $i$ and the public parameters $params$.

\ENSURE The client encrypted permission assignment list $L_{C_i}$ assigned to the client generated role $c^*_i (r)$.

\medskip

\STATE $c^*_i (r) \leftarrow$ call \textbf{ClientEnc} ($r$, $K_{u_i}$, $params$) \label{line:deploy-pa-cs-role}

\STATE $L_{C_i} \leftarrow \phi$ \label{line:deploy-pa-cs-init}

\FOR {each permission $(action, target)$ in $L$} \label{line:deploy-pa-cs-loop}

	\STATE $c^*_i (action) \leftarrow$ call \textbf{ClientEnc} ($action$, $K_{u_i}$, $params$) \label{line:deploy-pa-cs-action}
	
	\STATE $c^*_i (target) \leftarrow$ call \textbf{ClientEnc} ($target$, $K_{u_i}$, $params$) \label{line:deploy-pa-cs-target}
	
	\STATE $L_{C_i} \leftarrow L_{C_i} \cup (c^*_i (action), c^*_i (target))$ \label{line:deploy-pa-cs-update}

\ENDFOR

\RETURN ($c^*_i (r)$, $L_{C_i}$)

\end{algorithmic}
}
\end{algorithm}

% deploy: permission assignment: server side

\begin{algorithm}[htp]
{\algofontsize
\caption{\textbf{PermissionAssignment:ServerSide}}

\label{algo:deploy-permission-assignment-server-side}

\begin{algorithmic}[1]

\REQUIRE The client encrypted permission assignment list $L_{C_i}$ for client generated role $c^*_i (r)$ and identity $i$ of Admin User.

\ENSURE The server encrypted permission assignment list $L_{S}$ and the server generated role $c(r)$.

\medskip

\STATE $K_{s_i} \leftarrow KS[i]$ {\algofontsize \COMMENT{retrieve the server side key corresponding to Admin User $i$}} \label{line:deploy-pa-ss-ks}

\STATE $c(r) \leftarrow$ call \textbf{ServerReEnc} ($c^*_i (r)$, $K_{s_i}$) \label{line:deploy-pa-ss-role}

\STATE $L_{S} \leftarrow \phi$ \label{line:deploy-pa-ss-init}

\FOR {each client encrypted permission $(c^*_i (action), c^*_i (target))$ in list $L_{C_i}$} \label{line:deploy-pa-ss-loop}

	\STATE $c(action) \leftarrow$ call \textbf{ServerReEnc} ($c^*_i (action)$, $K_{s_i}$) \label{line:deploy-pa-ss-action}
	
	\STATE $c(target) \leftarrow$ call \textbf{ServerReEnc} ($c^*_i (target)$, $K_{s_i}$) \label{line:deploy-pa-ss-target}
	
	\STATE $L_{S} \leftarrow L_{S} \cup (c(action), c(target))$ \label{line:deploy-pa-ss-update}

\ENDFOR

\RETURN ($c(r)$, $L_{S}$)

\end{algorithmic}
}
\end{algorithm}

\emph{\textbf{Deployment of Permission Assignment Policies:}} 
An Admin User can assign permissions to a role. In order to deploy policies regarding permissions assignment to roles, an Admin User runs Algorithm \ref{algo:deploy-permission-assignment-client-side}. This algorithm takes as input a list of permissions $L$ to be assigned to role $r$, the client side key set $K_{u_i}$ corresponding to Admin User $i$ and the public parameters $params$ and outputs the client encrypted permission assignment list $L_{C_i}$ assigned to client generated role $c^*_i (r)$. First, it generates client encrypted role $c^*_i (r)$ by calling \textbf{ClientEnc} illustrated in Algorithm \ref{algo:client-enc} (Line \ref{line:deploy-pa-cs-role}). Next, it creates and initialises new list $L_{C_i}$ (Line \ref{line:deploy-pa-cs-init}). For each permission in $L$ (Line \ref{line:deploy-pa-cs-loop}), it generates the client encrypted action $c^*_i (action)$ (Line \ref{line:deploy-pa-cs-action}) and the client encrypted target $c^*_i (target)$ (Line \ref{line:deploy-pa-cs-target}) and updates $L_{C_i}$ by adding the client encrypted permission (Line \ref{line:deploy-pa-cs-update}). An Admin User sends the client encrypted permission list along with the client encrypted role to the Administration Point. 
The Administration Point runs another round of encryption by running Algorithm \ref{algo:deploy-permission-assignment-server-side}. This algorithm takes as input the client encrypted permission assignment list $L_{C_i}$ for client generated role $c^*_i (r)$ and identity $i$ of Admin User and outputs the server encrypted permission assignment list $L_{S}$ and the server generated role $c(r)$. First, it retrieves from the Key Store the server side key set $K_{s_i}$ corresponding to Admin User $i$ (Line \ref{line:deploy-pa-ss-ks}). Next, it generates the server encrypted role by calling \textbf{ServerReEnc} illustrated in Algorithm \ref{algo:server-re-enc} (Line \ref{line:deploy-pa-ss-role}). Then, it creates and initialises new list $L_{S}$ (Line \ref{line:deploy-pa-ss-init}). For each client encrypted role in $L_{C_i}$ (Line \ref{line:deploy-pa-ss-loop}), it generates the server encrypted action (Line \ref{line:deploy-pa-ss-action}) and the server encrypted target (Line \ref{line:deploy-pa-ss-target}) and updates $L_{S}$ by adding the server encryption permission (Line \ref{line:deploy-pa-ss-update}).

% deploy contextual condition: client side

\begin{algorithm}[htp]
{\algofontsize
\caption{\textbf{ContextualConditionDeployment:ClientSide}}

\label{algo:deploy-contextual-condition-client-side}

\begin{algorithmic}[1]

\REQUIRE The contextual condition $T$, the client side key set $K_{u_i}$ corresponding to Admin User $i$ and the public parameters $params$.

\ENSURE The client encrypted contextual condition $T_{C_i}$.

\medskip

\STATE $T_{C_i} \leftarrow T$ \label{line:deploy-cc-cs-copy}

\FOR {each leaf node $e$ in $T_{C_i}$} \label{line:deploy-cc-cs-loop}

	\STATE $c^*_i (e) \leftarrow$ call \textbf{ClientEnc} ($r$, $K_{u_i}$, $params$) \label{line:deploy-cc-cs-call}
	
	\STATE replace $e$ of $T_{C_i}$ with $c^*_i (e)$ \label{line:deploy-cc-cs-replace}

\ENDFOR

\RETURN $T_{C_i}$

\end{algorithmic}
}
\end{algorithm}

% deploy contextual condition: server side

\begin{algorithm}[htp]
{\algofontsize
\caption{\textbf{ContextualConditionDeployment:ServerSide}}

\label{algo:deploy-contextual-condition-server-side}

\begin{algorithmic}[1]

\REQUIRE The client encrypted contextual condition $T_{C_i}$ and identity of Admin User $i$.

\ENSURE The server encrypted contextual condition $T_{S}$

\medskip

\STATE $K_{s_i} \leftarrow KS[i]$ {\algofontsize \COMMENT{retrieve the server side key corresponding to Admin User $i$}} \label{line:deploy-cc-ss-ks}

\STATE $T_{S} \leftarrow T_{C_i}$ \label{line:deploy-cc-ss-copy}

\FOR {each client encrypted leaf node $c^*_i (e)$ in $T_{S}$} \label{line:deploy-cc-ss-loop}

	\STATE $c(e) \leftarrow$ call \textbf{ServerReEnc} ($c^*_i (e)$, $K_{s_i}$) \label{line:deploy-cc-ss-call}
	
	\STATE replace $c^*_i (e)$ of $T_{S}$ with $c(e)$ \label{line:deploy-cc-ss-replace}

\ENDFOR

\RETURN $T_{S}$

\end{algorithmic}
}
\end{algorithm}

\emph{\textbf{Deployment of Contextual Conditions:}}
The contextual condition (part of role assignment and permission assignment policies) can be deployed in two steps. In the first step, an Admin User performs a first round of encryption by running Algorithm \ref{algo:deploy-contextual-condition-client-side}. This algorithm takes as input the contextual condition $T$, the client side key set $K_{u_i}$ corresponding to Admin User $i$ and the public parameters $params$ and outputs the client encrypted contextual condition $T_{C_i}$. First, it copies $T$ to $T_{C_i}$ (Line \ref{line:deploy-cc-cs-copy}). For each leaf node in $T_{C_i}$ (Line \ref{line:deploy-cc-cs-loop}), it generates the client encrypted element by calling \textbf{ClientEnc} illustrated in Algorithm \ref{algo:client-enc} (Line \ref{line:deploy-cc-cs-call}) and then updates $T_{C_i}$ by replacing element $e$ with the client encrypted element $c^*_i (e)$ (Line \ref{line:deploy-cc-cs-replace}). An Admin User sends the client encrypted contextual condition to the Administration Point.
In the second step, the Administration Point performs another round of encryption by running Algorithm \ref{algo:deploy-contextual-condition-server-side}. This algorithm takes as input the client encrypted contextual condition $T_{C_i}$ and identity of Admin User $i$ and outputs the server encrypted contextual condition $T_{S}$. First, it retrieves from the Key Store the server side key $K_{s_i}$ corresponding to Admin User $i$ (Line \ref{line:deploy-cc-ss-ks}). Next, it copies $T_{C_i}$ to $T_{S}$ (Line \ref{line:deploy-cc-ss-copy}). For each each client encrypted leaf node in $T_{S}$ (Line \ref{line:deploy-cc-ss-loop}), it generates the server encrypted element by calling \textbf{ServerReEnc} illustrated in Algorithm \ref{algo:server-re-enc} (Line \ref{line:deploy-cc-ss-call}). Then, it replaces the client encrypted element $c^*_i (e)$ of $T_{S}$ with the server encrypted element $c(e)$ (Line \ref{line:deploy-cc-ss-replace}).

% deploy role hierarchy: client side

\begin{algorithm}[htp]
{\algofontsize
\caption{\textbf{RoleHierarchyDeployment:ClientSide}}

\label{algo:deploy-role-hierarchy-client-side}

\begin{algorithmic}[1]

\REQUIRE The role hierarchy graph $G$, the client side key set $K_{u_i}$ corresponding to Admin User $i$ and the public parameters $params$.

\ENSURE The client generated role hierarchy graph $G_{C_i}$.

\medskip

\STATE $G_{C_i} \leftarrow G$ \label{line:deploy-rh-cs-copy}

\FOR {each node $r$ in $G_{C_i}$} \label{line:deploy-rh-cs-loop}

	\STATE $c^*_i (r) \leftarrow$ call \textbf{ClientEnc} ($r$, $K_{u_i}$, $params$) \label{line:deploy-rh-cs-enc}
	\STATE $td^*_i (r) \leftarrow$ call \textbf{ClientTD} ($r$, $K_{u_i}$, $params$) {\algofontsize \COMMENT{see Algorithm \ref{algo:client-td}}} \label{line:deploy-rh-cs-td}
	\STATE replace $r$ of $G_{C_i}$ with $(c^*_i (r), td^*_i (r))$ \label{line:deploy-rh-cs-replace}

\ENDFOR

\RETURN $G_{C_i}$

\end{algorithmic}
}
\end{algorithm}

% deploy role hierarchy: server side

\begin{algorithm}[htp]
{\algofontsize
\caption{\textbf{RoleHierarchyDeployment:ServerSide}}

\label{algo:deploy-role-hierarchy-server-side}

\begin{algorithmic}[1]

\REQUIRE The client generated role hierarchy graph $G_{C_i}$ and identity of Admin User $i$.

\ENSURE The server generated role hierarchy graph $G_{S}$

\medskip

\STATE $K_{s_i} \leftarrow KS[i]$ {\algofontsize \COMMENT{retrieve the server side key corresponding to Admin User $i$}} \label{line:deploy-rh-ss-ks}

\STATE $G_{S} \leftarrow G_{C_i}$ \label{line:deploy-rh-ss-copy}

\FOR {each client generated node $(c^*_i (r), td^*_i (r))$ in $G_{S}$} \label{line:deploy-rh-ss-loop}

	\STATE $c(r) \leftarrow$ call \textbf{ServerReEnc} ($c^*_i (r)$, $K_{s_i}$) \label{line:deploy-rh-ss-enc}
	
	\STATE $td(r) \leftarrow$ call \textbf{ServerTD} ($td^*_i (r)$, $K_{s_i}$) {\algofontsize \COMMENT{see Algorithm \ref{algo:server-td}}} \label{line:deploy-rh-ss-td}
	
	\STATE replace $(c^*_i (r), td^*_i (r))$ of $G_{S}$ with $(c(r), td(r))$ \label{line:deploy-rh-ss-replace}

\ENDFOR

\RETURN $G_{S}$

\end{algorithmic}
}
\end{algorithm}

\emph{\textbf{Deployment of Role Hierarchy Graph:}} 
We know that a derived role inherits all permissions from its base role. In case if requested permissions are not assigned to the Requester's role, the PDP may need to traverse in the role hierarchy graph to find base roles corresponding to the Requester's role and then PDP verifies if any base role can fulfil requested permissions. For this purpose, the PDP needs a trapdoor of each base role so that it can match this trapdoor against roles in the Permission Repository. Therefore, a role hierarchy graph stores a role trapdoor along with each encrypted role. The deployment of role hierarchy graph takes place in two steps. In the first step, an Admin User runs Algorithm \ref{algo:deploy-role-hierarchy-client-side}. This algorithm takes as input the role hierarchy graph $G$, the client side key set $K_{u_i}$ corresponding to Admin User $i$ and the public parameters $params$ and outputs the client generated role hierarchy graph $G_{C_i}$. First, it copies $G$ to $G_{C_i}$  (Line \ref{line:deploy-rh-cs-copy}). For each node $r$ in $G_{C_i}$ (Line \ref{line:deploy-rh-cs-loop}), it generates the client encrypted role by calling \textbf{ClientEnc} illustrated in Algorithm \ref{algo:client-enc} (Line \ref{line:deploy-rh-cs-enc}) and the client trapdoor by calling \textbf{ClientTD} (Line \ref{line:deploy-rh-cs-td}) illustrated in Algorithm \ref{algo:client-td} that is explained later in this section. Next, it replaces $r$ of $G_{C_i}$ with the client encrypted role and the client generated trapdoor (Line \ref{line:deploy-rh-cs-replace}). An Admin User sends the client generated role hierarchy graph to the Administration Point. 
In the second step, the Administration Point runs Algorithm \ref{algo:deploy-role-hierarchy-server-side}. This algorithm takes as input the client generated role hierarchy graph $G_{C_i}$ and identity of Admin User $i$ and outputs the server generated role hierarchy graph $G_{S}$. First, it retrieves from the Key Store the server side key $K_{s_i}$ corresponding to Admin User $i$ (Line \ref{line:deploy-rh-ss-ks}). Next, it copies $G_{C_i}$ to $G_{S}$ (Line \ref{line:deploy-rh-ss-copy}). For each client generated node (Line \ref{line:deploy-rh-ss-loop}), it generates the server encrypted role by calling \textbf{ServerReEnc} illustrated in Algorithm \ref{algo:server-re-enc} (Line \ref{line:deploy-rh-ss-enc}) and the server trapdoor by calling \textbf{ServerTD} (Line \ref{line:deploy-rh-ss-td}) illustrated in Algorithm \ref{algo:server-td} that is explained later in this section and then updates $G_{S}$ by replacing the client generated node with the server generated node (Line \ref{line:deploy-rh-ss-replace}).

% aux method: client td

\begin{algorithm}[h]
{\algofontsize
\caption{\textbf{ClientTD}}

\label{algo:client-td}

\begin{algorithmic}[1]

\REQUIRE Element $e$, the client side key set $K_{u_i}$ corresponding to user $i$ and the public parameters $params$.

\ENSURE The client generated trapdoor $td^*_i (e)$.

\medskip

\STATE Choose a random $r_{e} \in \mathbb{Z}^*_q$ \label{line:c-td-choose}
\STATE ${\sigma}_{e} \leftarrow f_s (e)$ \label{line:c-td-sigma}
\STATE $t_1 \leftarrow g^{-r_{e}} g^{{\sigma}_{e}}$ \label{line:c-td-t1}
\STATE $t_2 \leftarrow h^{r_{e}} g^{-x_{i1}r_{e}} g^{x_{i1}{\sigma}_{e}} = g^{x_{i2}r_{e}} g^{x_{i1}{\sigma}_{e}}$ \label{line:c-td-t2}

\STATE $td^*_i (e) \leftarrow (t_1, t_2)$ \label{line:c-td-td}

\RETURN $td^*_i (e)$

\end{algorithmic}
}
\end{algorithm}

% aux method: server td

% server td
\begin{algorithm}[htp]
{\algofontsize
\caption{\textbf{ServerTD}}

\label{algo:server-td}

\begin{algorithmic}[1]

\REQUIRE The client generated trapdoor $td^*_i (e)$ and the server side key set $K_{s_i}$ corresponding to user $i$.

\ENSURE The server generated trapdoor $td(e)$.

\medskip

\STATE $td(e) \leftarrow t_1^{x_{i2}} . t_2 = g^{x{\sigma}_{e}}$ \label{line:s-td-calculate}

\RETURN $td(e)$

\end{algorithmic}
}
\end{algorithm}

\begin{figure}
\centering
% left bottom right top
\includegraphics[trim=70mm 50mm 30mm 130mm,clip,width=.45\textwidth]{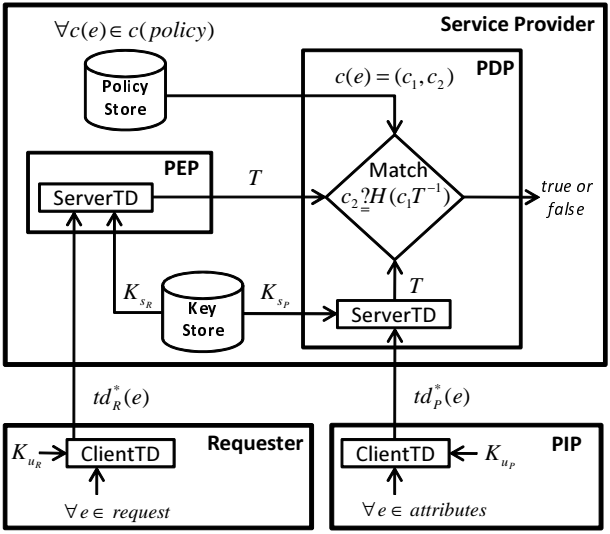}
\caption{Policy evaluation phase}
\label{fig:doctor}
\end{figure}

% aux method: match elements

\begin{algorithm}[h]
{\algofontsize
\caption{\textbf{Match}}

\label{algo:match}

\begin{algorithmic}[1]

\REQUIRE The server encrypted element $c(e) = (c_1, c_2)$ and the server generated trapdoor $td(e) = T$.

\ENSURE $\mathit{true}$ or $\mathit{false}$

\medskip

\IF {$c_2 \stackrel{?}{=} H(c_1 . T^{-1})$} \label{line:match-condition}

	\RETURN $\mathit{true}$ \label{line:match-true}
	
\ELSE

	\RETURN $\mathit{false}$ \label{line:match-false}
	
\ENDIF

\end{algorithmic}
}
\end{algorithm}

\subsection{Policy Evaluation Phase}

The policy evaluation phase is executed when a Requester makes a request either $\mathit{ACT}$ or $\mathit{REQ}$. In this phase, a Requester sends client generated trapdoors (using Algorithm \ref{algo:client-td}) of a request to the PEP. The PEP converts client generated trapdoors into server generated trapdoors (using Algorithm \ref{algo:server-td}) and sends them to the PDP. The PDP matches server encrypted trapdoors of the request with server encrypted elements of the policy (using Algorithm \ref{algo:match}). Optionally, the PDP may require contextual information in order to evaluate contextual conditions. The PIP sends client generated trapdoors of contextual information to the PDP. The PDP converts client generated trapdoors into server generated trapdoors and then evaluates contextual conditions based on contextual information. Finally, the PDP returns either $\mathit{true}$ or $\mathit{false}$ as shown in Figure \ref{fig:doctor}. In the following, we describe how we generate trapdoors and perform the match. 

For calculating client generated trapdoors of a request (or contextual information), a Requester (or the PIP) runs \textbf{ClientTD} illustrated in Algorithm \ref{algo:client-td}. \textbf{ClientTD} takes as input each element $e$ of the request, the client side key set $K_{u_i}$ corresponding to user $i$ and the public parameters $params$ and outputs the client generated trapdoor $td^*_i (e)$. First, it choose randomly $r_{e} \in \mathbb{Z}^*_q$ (Line \ref{line:c-td-choose}). Next, it calculates ${\sigma}_{e}$ as $f_s (e)$ (Line \ref{line:c-td-sigma}). Then it calculates $t_1$ and $t_2$ as $g^{-r_{e}} g^{{\sigma}_{e}}$ (Line \ref{line:c-td-t1}) and $h^{r_{e}} g^{-x_{i1}r_{e}} g^{x_{i1}{\sigma}_{e}} = g^{x_{i2}r_{e}} g^{x_{i1}{\sigma}_{e}}$ (Line \ref{line:c-td-t2}), respectively. Both $t_1$ and $t_2$ form $td^*_i (e)$ (Line \ref{line:c-td-td}). A Requester sends client generated trapdoors of the request to the PEP.
The PEP receives client generated trapdoors and runs \textbf{ServerTD} illustrated in Algorithm \ref{algo:server-td} for calculating server generated trapdoors. \textbf{ServerTD} takes as input the client generated trapdoor $td^*_i (e)$ and the server side key set $K_{s_i}$ corresponding to user $i$ and outputs the server generated trapdoor $td(e)$. It calculates $td(e)$ as  $t_1^{x_{i2}} . t_2 = g^{x{\sigma}_{e}}$ (Line \ref{line:s-td-calculate}).

In order to match a server encrypted element of a policy with a server generated trapdoor of a request, the PDP runs \textbf{Match} illustrated in Algorithm \ref{algo:match}. \textbf{Match} takes as input the server encrypted element $c(e) = (c_1, c_2)$ and the server generated trapdoor $td(e) = T$ and returns either $\mathit{true}$ or $\mathit{false}$. It checks the condition $c_2 \stackrel{?}{=} H(c_1 . T^{-1})$ (Line \ref{line:match-condition}). If the condition holds, it returns $\mathit{true}$ (Line \ref{line:match-true}) indicating that the match is successful. Otherwise, it returns $\mathit{false}$ (Line \ref{line:match-false}). 

In the following, we describe how to evaluate (parts of) policies including role assignment, permission assignment, contextual conditions and role hierarchy graph. For the evaluation of each (part of) policy, we follow general strategy as already described in this section and also illustrated in Figure \ref{fig:doctor}.

% search role in repository or session

\begin{algorithm}[htp]
{\algofontsize
\caption{\textbf{SearchRole}}

\label{algo:search-role}

\begin{algorithmic}[1]

\REQUIRE The client generated trapdoor of role $td^*_i (r)$ and the server encrypted role assignment list (or list of active roles in session) $L_{S}$ for Requester $i$

\ENSURE $\mathit{true}$ or $\mathit{false}$

\medskip

\STATE $K_{s_i} \leftarrow KS[i]$ {\algofontsize \COMMENT{retrieve the server side key corresponding to Requester $i$}} \label{line:search-role-ks}

\STATE $td(r) \leftarrow$ call \textbf{ServerTD} ($td^*_i (r)$, $K_{s_i}$) \label{line:search-role-td}

\FOR {each server encrypted role $c(r)$ in $L_{S}$} \label{line:search-role-loop}

	\STATE $match \leftarrow$ call \textbf{Match} ($c(r)$, $td(r)$) {\algofontsize \COMMENT{see Algorithm \ref{algo:match}}} \label{line:search-role-call}

	\IF {$match \stackrel{?}{=} true$} \label{line:search-role-match}
		\RETURN $\mathit{true}$ \label{line:search-role-true}
	\ENDIF

\ENDFOR

\RETURN $\mathit{false}$ \label{line:search-role-false}

\end{algorithmic}
}
\end{algorithm}

\emph{\textbf{Searching a Role:}} 
A Requester can make a role activation request $\mathit{ACT}$ and sends it to the SP. In order to grant $\mathit{ACT}$, the SP runs \textbf{SearchRole} illustrated in Algorithm \ref{algo:search-role}. This algorithm takes as input the client generated trapdoor of role $td^*_i (r)$ and the server encrypted role assignment list $L_{S}$ for Requester $i$. First, it retrieves from the Key Store the server side key $K_{s_i}$ corresponding to Requester $i$ (Line \ref{line:search-role-ks}). Next, it calculates the server generated trapdoor $td(r)$ by calling Algorithm \ref{algo:server-td} (Line \ref{line:search-role-td}). For each server encrypted role $c(r)$ in $L_{S}$ (Line \ref{line:search-role-loop}), it performs matching against $td(r)$ by calling Algorithm \ref{algo:match} (Line \ref{line:search-role-call}). If any match is successful (Line \ref{line:search-role-match}), it returns $\mathit{true}$ (Line \ref{line:search-role-true}), meaning that $\mathit{ACT}$ is granted. Otherwise, it returns $\mathit{false}$ (Line \ref{line:search-role-false}).

After $\mathit{ACT}$ is granted, the PEP updates Session by adding in the Active Roles repository the server generated trapdoor of role. Once a Requester is active in a role, she can make an access request $\mathit{REQ}$. Before granting $\mathit{REQ}$, the SP checks if the Requester is already in the role in $\mathit{REQ}$. For this purpose, the SP runs Algorithm \ref{algo:search-role}, where $L_{S}$ shows a list of active roles in the session. Furthermore, the PDP also runs Algorithm \ref{algo:search-role} for searching the role in $\mathit{REQ}$ in the Permission Repository with a slight modification of ignoring the server trapdoor generation (in Line \ref{line:search-role-td}) as it is already generated when the role of $\mathit{REQ}$ is searched in the session.

% search permissions

\begin{algorithm}[htp]
{\algofontsize
\caption{\textbf{SearchPermission}}

\label{algo:search-permission}

\begin{algorithmic}[1]

\REQUIRE The client generated trapdoor of permission ($td^*_i (action)$, $td^*_i (target)$ and the server encrypted permission assignment list $L_{S}$ for Requester $i$

\ENSURE $\mathit{true}$ or $\mathit{false}$

\medskip

\STATE $K_{s_i} \leftarrow KS[i]$ {\algofontsize \COMMENT{retrieve the server side key corresponding to Requester $i$}} \label{line:search-permission-ks}

\STATE $td(action) \leftarrow$ call \textbf{ServerTD} ($td^*_i (action)$, $K_{s_i}$) \label{line:search-permission-td-a}

\STATE $td(target) \leftarrow$ call \textbf{ServerTD} ($td^*_i (target)$, $K_{s_i}$) \label{line:search-permission-td-t}

\FOR {each server encrypted permission $(c(action), c(target))$ in $L_{S}$} \label{line:search-permission-loop}

	\STATE $match_{action} \leftarrow$ call \textbf{Match} ($c(action)$, $td(action)$) \label{line:search-permission-match-a}
	
	\STATE $match_{target} \leftarrow$ call \textbf{Match} ($c(target)$, $td(target)$) \label{line:search-permission-match-t}

	\IF {$match_{action} \stackrel{?}{=} true$ and $match_{target} \stackrel{?}{=} true$} \label{line:search-permission-check-match}
		\RETURN $\mathit{true}$ \label{line:search-permission-true}
	\ENDIF

\ENDFOR

\RETURN $\mathit{false}$ \label{line:search-permission-false}

\end{algorithmic}
}
\end{algorithm}

\emph{\textbf{Searching a Permission:}} A Requester can send $\mathit{REQ}$ for executing certain permissions. The PEP on the SP checks if the Requester is active in the role indicated in $\mathit{REQ}$ and then the searches that role in the Permission Repository by running Algorithm \ref{algo:search-role}. After a role is matched in the Permission Repository, the PEP searches the permission in $\mathit{REQ}$ by running Algorithm \ref{algo:search-permission}. This algorithm takes as input the client generated trapdoor of permission ($td^*_i (action)$, $td^*_i (target)$ and the server encrypted permission assignment list $L_{S}$ for Requester $i$ and returns either $\mathit{true}$ or $\mathit{false}$. First, it retrieves from the Key Store from the Key Store the server side key $K_{s_i}$ corresponding to Requester $i$ (Line \ref{line:search-permission-ks}). Next, it calculates server generated trapdoors of both action (Line \ref{line:search-permission-td-a}) and target (Line \ref{line:search-permission-td-t}) by calling Algorithm \ref{algo:server-td}. For each server encrypted permission $(c(action), c(target))$ in $L_{S}$ (Line \ref{line:search-permission-loop}), it matches the server encrypted action with the server generated action (Line \ref{line:search-permission-match-a}) and the server encrypted target with the server generated taret (Line \ref{line:search-permission-match-t}), respectively, by calling Algorithm \ref{algo:match}. If both matches are successful (Line \ref{line:search-permission-check-match}) for any permission $(c(action), c(target))$ in $L_{S}$, it returns $\mathit{true}$ (Line \ref{line:search-permission-true}). Otherwise, it returns $\mathit{false}$ (Line \ref{line:search-permission-false}).

% contextual condition: client side: request

\begin{algorithm}[htp]
{\algofontsize
\caption{\textbf{ContextualConditionRequest}}

\label{algo:request-contextual-condition}

\begin{algorithmic}[1]

\REQUIRE List of attributes contextual attributes $L$, the client side key set $K_{u_i}$ corresponding to Requester $i$ and the public parameters $params$.

\ENSURE The client generated list of trapdoors of contextual attributes $L_{C_i}$.

\medskip

\STATE $L_{C_i} \leftarrow \phi$ \label{request-cc-init}

\FOR {each attribute $e$ in $L$} \label{request-cc-loop}

	\STATE $td^*_i (e) \leftarrow$ call \textbf{ClientTD} ($r$, $K_{u_i}$, $params$) \label{request-cc-td}
	
	\STATE $L_{C_i} \leftarrow L_{C_i} \cup td^*_i (e)$ \label{request-cc-update}

\ENDFOR

\RETURN $T_{C_i}$

\end{algorithmic}
}
\end{algorithm}

\emph{\textbf{Generating Contextual Attributes:}} 
The PIP runs \textbf{ContextualAttributesRequest} illustrated in Algorithm \ref{algo:request-contextual-condition} to calculate client generated trapdoors of contextual information. \textbf{ContextualAttributesRequest} takes as input a list of contextual attributes $L$, the client side key set $K_{u_i}$ corresponding to Requester $i$ and the public parameters $params$ and outputs the client generated list of trapdoors of contextual attributes $L_{C_i}$. First, it creates and initialises new list $L_{C_i}$ (Line \ref{request-cc-init}). For each attribute $e$ in $L$ (Line \ref{request-cc-loop}), it calculates the client generated trapdoor $td^*_i (e)$ by calling Algorithm \ref{algo:client-td} (Line \ref{request-cc-td}) and adds $td^*_i (e)$ in $L_{C_i}$ (Line \ref{request-cc-update}).

% aux EvaluateTree method

\begin{algorithm}[htp]

{\algofontsize
\caption{\textbf{EvaluateTree}}

\label{algo:evaluate-tree}

\begin{algorithmic}[1]

\REQUIRE Node $n$ and tree $T$.

\ENSURE $\mathit{true}$ or $\mathit{false}$.

\medskip

\IF {$n.decision \neq null$} \label{line:evaluate-tree-if-null}
	\RETURN $n.decision$ \label{line:evaluate-tree-decision-not-null}
\ENDIF

\FOR {each child $c$ of $n$ in tree $T$} \label{line:evaluate-tree-child-loop}
	
	\STATE call \textbf{EvaluateTree} ($c$, $T$) {\algofontsize \COMMENT{recursive call}} \label{line:evaluate-tree-call}
	
\ENDFOR

\STATE $t \leftarrow 0$ \label{line:evaluate-tree-init-t}

\STATE $m \leftarrow 0$ \label{line:evaluate-tree-init-m}

\FOR {each child $c$ of $n$ in tree $T$} \label{line:evaluate-tree-count-loop}

	\STATE $t \leftarrow t + 1$ \label{line:evaluate-tree-inc-t}
	
	\IF {$c.decision \stackrel{?}{=} true$} \label{line:evaluate-tree-find-decision}
		\STATE $m \leftarrow m + 1$ \label{line:evaluate-tree-inc-m}
	\ENDIF

\ENDFOR

\IF {($n.gate \stackrel{?}{=} AND$ and $m \stackrel{?}{=} t$) or ($n.gate \stackrel{?}{=} OR$ and $m \geq 1$)} \label{line:evaluate-tree-find-gate}
	\STATE $n.decision \leftarrow true$ \label{line:evaluate-tree-set-decision-true}
\ELSE
	\STATE $n.decision \leftarrow false$ \label{line:evaluate-tree-set-decision-false}
\ENDIF

\RETURN $n.decision$ \label{line:evaluate-tree-return-decision}

\end{algorithmic}

}

\end{algorithm}

% Moving ContextualConditionEvaluation from here: AAA

\emph{\textbf{Evaluating Contextual Conditions:}}
For evaluating any contextual condition, the PDP runs \textbf{ContextualConditionEvaluation} illustrated in Algorithm \ref{algo:match-contextual-condition}. This algorithm takes as input the client generated list of trapdoors of contextual attributes $L_{C_i}$, the server encrypted contextual condition $T_{S}$ and identity of Requester $i$ and returns either $\mathit{true}$ or $\mathit{false}$. First, it retrieves from the Key Store the server side key $K_{s_i}$ corresponding to Requester $i$ (Line \ref{line:cc-match-ks}). Next, it creates and initialises a new list $L_{S}$ (Line \ref{line:cc-match-init-list}). For each client generated trapdoor $td^*_i (e)$ in $L_{C_i}$ (Line \ref{line:cc-match-loop-s-td}), it calculates the server generated trapdoor $td(e)$ by calling Algorithm \ref{algo:server-td} (Line \ref{line:cc-match-call-s-td}) and adds $td(e)$ in $L_{S}$ (Line \ref{line:cc-match-update-list}). Next, it copies $T_{S}$ to $\mathit{TREE}$ (Line \ref{line:cc-match-copy-tree}) and adds decision field to each node in $\mathit{TREE}$ (Line \ref{line:cc-match-add-field}). For each node $n$ in $\mathit{TREE}$ (Line \ref{line:cc-match-loop-tree-init}), it initialises $n.decision$ as $null$ (Line \ref{line:cc-match-init-field}). For each leaf node $n$ in $\mathit{TREE}$ (Line \ref{line:cc-match-loop-tree-match}), it checks if any server generated trapdoor $td(e)$ in $L_{S}$ (Line \ref{line:cc-match-loop-list-match}) matches with it by calling Algorithm \ref{algo:match} (Line \ref{line:cc-match-decision}). Next, it evaluates non-leaf nodes of $\mathit{TREE}$ by running Algorithm \ref{algo:evaluate-tree} (Line \ref{line:cc-match-call-eval-tree}). Finally, it returns either $\mathit{true}$ or $\mathit{false}$ depending upon the evaluation of $\mathit{TREE}$ (Line \ref{line:cc-match-return-decision}). 

% moved here: AAA
% contextual condition: request and then search/matching

\begin{algorithm}[htp]
{\algofontsize
\caption{\textbf{ContextualConditionEvaluation}}

\label{algo:match-contextual-condition}

\begin{algorithmic}[1]

\REQUIRE The client generated list of trapdoors of contextual attributes $L_{C_i}$, the server encrypted contextual condition $T_{S}$ and identity of Requester $i$.

\ENSURE $\mathit{true}$ or $\mathit{false}$

\medskip

\STATE $K_{s_i} \leftarrow KS[i]$ {\algofontsize \COMMENT{retrieve the server side key corresponding to Requester $i$}} \label{line:cc-match-ks}

\STATE $L_{S} \leftarrow \phi$ \label{line:cc-match-init-list}

\FOR {each client generated trapdoor $td^*_i (e)$ in $L_{C_i}$} \label{line:cc-match-loop-s-td}
	
	\STATE $td(e) \leftarrow$ call \textbf{ServerTD} ($td^*_i (e)$, $K_{s_i}$) \label{line:cc-match-call-s-td}
	
	\STATE $L_{S} \leftarrow L_{S} \cup td^*_i (e)$ \label{line:cc-match-update-list}
	
\ENDFOR

\STATE $TREE \leftarrow T_{S}$ \label{line:cc-match-copy-tree}

\STATE Add decision field to each node in $\mathit{TREE}$ \label{line:cc-match-add-field}

\FOR {each node $n$ in $\mathit{TREE}$} \label{line:cc-match-loop-tree-init}

	\STATE $n.decision \leftarrow null$ \label{line:cc-match-init-field}

\ENDFOR

\FOR {each leaf node $n$ in $\mathit{TREE}$}	\label{line:cc-match-loop-tree-match}
	
	\FOR {each server generated trapdoor $td(e)$ in $L_{S}$} \label{line:cc-match-loop-list-match}
	
		\STATE $n.decision \leftarrow$ call \textbf{Match} ($n.c(e)$, $td(e)$) \label{line:cc-match-decision}

		\IF {$n.decision \stackrel{?}{=} true$} \label{line:cc-match-if-success}
			\RETURN $break;$ \label{line:cc-match-stop}
		\ENDIF  
	
	\ENDFOR

\ENDFOR

\STATE call \textbf{EvaluateTree} ($TREE.root$, $\mathit{TREE}$) {\algofontsize \COMMENT{see Algorithm \ref{algo:evaluate-tree}}} \label{line:cc-match-call-eval-tree}

\RETURN $TREE.root.decision$ \label{line:cc-match-return-decision}

\end{algorithmic}
}
\end{algorithm}

\textbf{EvaluateTree} evaluates a tree containing AND and OR gates. It takes as input root node $n$ and tree $T$ and returns either $\mathit{true}$ or $\mathit{false}$. First, it checks if the decision for $n$ is already made (Line \ref{line:evaluate-tree-if-null}). If so, it returns the decision (Line \ref{line:evaluate-tree-decision-not-null}). For each child $c$ of $n$ in tree $T$ (Line \ref{line:evaluate-tree-child-loop}), it recursively calls \textbf{EvaluateTree} (Line \ref{line:evaluate-tree-call}). Next, it creates and initialises $t$ (Line \ref{line:evaluate-tree-init-t}) and $m$ (Line \ref{line:evaluate-tree-init-m}) indicating total children of $n$ and a count of matched children, respectively. For each child $c$ of $n$ in tree $T$ (Line \ref{line:evaluate-tree-count-loop}), it counts total children (Line \ref{line:evaluate-tree-inc-t}) and matched children by checking made decisions (Line \ref{line:evaluate-tree-inc-m}). Next, it checks if non-leaf node is AND and all children are matched or non-leaf node is OR and at least one child is matched (Line \ref{line:evaluate-tree-find-gate}). If so, it is set as $\mathit{true}$ (Line \ref{line:evaluate-tree-set-decision-true}) and $\mathit{false}$ (Line \ref{line:evaluate-tree-set-decision-false}) otherwise.

% search role hierarchy

\begin{algorithm}[htp]
{\algofontsize
\caption{\textbf{SearchRoleHierarchyGraph}}

\label{algo:search-role-hierarchy-graph}

\begin{algorithmic}[1]

\REQUIRE The server generated trapdoor of role $td(r)$ and the server generated role hierarchy graph $G_{S}$

\ENSURE $\mathit{true}$ or $\mathit{false}$

\medskip

\FOR {each server encrypted role $c(r)$ in $G_{S}$} \label{line:search-rh-loop}

	\STATE $match \leftarrow$ call \textbf{Match} ($c(r)$, $td(r)$) \label{line:search-rh-call}

	\IF {$match \stackrel{?}{=} true$} \label{line:search-rh-match}
		\RETURN $\mathit{true}$ \label{line:search-rh-true}
	\ENDIF

\ENDFOR

\RETURN $\mathit{false}$ \label{line:search-rh-false}

\end{algorithmic}
}
\end{algorithm}

\emph{\textbf{Searching Roles in Role Hierarchy Graph:}}
The PDP may need to search base roles of one in $\mathit{REQ}$ since a derived role inherits all permissions from its base role. The PDP runs \textbf{SearchRoleHierarchyGraph} illustrated in Algorithm \ref{algo:search-role-hierarchy-graph} to find base roles from the encrypted role hierarchy graph. This algorithm takes as input the server generated trapdoor of role $td(r)$ and the server generated role hierarchy graph $G_{S}$ and returns $\mathit{true}$ if any base role is found and $\mathit{false}$ otherwise. For each server encrypted role $c(r)$ in $G_{S}$ (Line \ref{line:search-rh-loop}), it checks if $td(r)$ matches with any $c(r)$ by calling Algorithm \ref{algo:match} (Line \ref{line:search-rh-call}). If any match is found (Line \ref{line:search-rh-match}), it returns $\mathit{true}$ (Line \ref{line:search-rh-true}). Otherwise, it returns $\mathit{false}$ (Line \ref{line:search-rh-false}).

% user revocation

\begin{algorithm}[htp]
{\algofontsize
\caption{\textbf{UserRevocation}}

\label{algo:user-revocation}

\begin{algorithmic}[1]

\REQUIRE The user identity $i$.

\ENSURE $\mathit{true}$ or $\mathit{false}$.

\medskip

\IF {$exits(KS[i]) \stackrel{?}{=} false$} \label{line:user-revocation-exist}

	\RETURN $\mathit{false}$ \label{line:user-revocation-false}

\ENDIF

\STATE $K_{s_i} \leftarrow KS[i]$ \label{line:user-revocation-get}
\STATE $KS \leftarrow KS \backslash K_{s_i}$ \label{line:user-revocation-remove}

\RETURN $\mathit{true}$ \label{line:user-revocation-true}

\end{algorithmic}
}
\end{algorithm}

\subsection{Revocation Phase}
In this phase, the PEP can remove a compromised user from the system. In order to remove a user, the PEP runs \textbf{UserRevocation} illustrated in Algorithm \ref{algo:user-revocation}. This algorithm takes as input the user identity $i$ and returns either $\mathit{true}$ (indicating that the user has been removed successfully) or false (indicating that the user does not exist in the system). First, it checks if the given user exists by checking the Key Store. If no, it returns $\mathit{false}$ (Line \ref{line:user-revocation-false}). Otherwise, it retrieves from the Key Store the server side key set $K_{s_i}$ corresponding to user $i$ (Line \ref{line:user-revocation-get}), removes $K_{s_i}$ from the Key Store (Line \ref{line:user-revocation-remove}) and returns $\mathit{true}$ (Line \ref{line:user-revocation-true}).

\section{Security Analysis}
\label{sec:security-analysis} 
In this section, we analyse the security of the policy deployment phase that includes Role Assignment (RA) encryption (Algorithms \ref{algo:deploy-role-assignment-client-side} and \ref{algo:deploy-role-assignment-server-side}), Permission Assignment (PA) encryption (Algorithms \ref{algo:deploy-permission-assignment-client-side} and \ref{algo:deploy-permission-assignment-server-side}), Contextual Condition (CC) encryption (Algorithms \ref{algo:deploy-contextual-condition-client-side} and \ref{algo:deploy-contextual-condition-server-side}), and Role Hierarchy (RH) encryption (Algorithms \ref{algo:deploy-role-hierarchy-client-side} and \ref{algo:deploy-role-hierarchy-server-side}). We then analyse the security of the policy evaluation phase that include Search Role (SR) (Algorithms \ref{algo:client-td} and \ref{algo:search-role}), Search Permission (Algorithms \ref{algo:client-td} and \ref{algo:search-permission}), Contextual Condition Evaluation (Algorithms \ref{algo:request-contextual-condition} and \ref{algo:match-contextual-condition}) and Search Role Hierarchy (Algorithms \ref{algo:client-td}, \ref{algo:server-td} and \ref{algo:search-role-hierarchy-graph}). 

We first define some basic concepts on which we build our security proofs.

\subsection{Preliminaries}
In general, a scheme is considered secure if no adversary can break the scheme with probability significantly greater than random guessing. The adversary's advantage in breaking the scheme should be a negligible function of the security parameter.

\begin{definition}[Negligible Function]
A function $f$ is negligible if for each polynomial $p()$ there exists $N$ such that for all integers $n > N$ it holds that $f(n)<\frac{1}{p(n)}$.
\end{definition}

We consider a realistic adversary that is computationally bounded and show that our scheme is secure against such an adversary. We model the adversary as a randomised algorithm that runs in polynomial time and show that the success probability of any such adversary is negligible. An algorithm that is randomised and runs in polynomial time is
called a Probabilistic Polynomial Time (PPT) algorithm.

Our scheme relies on the existence of a pseudorandom function $f$. Intuitively, the output a pseudorandom function cannot be distinguished by a realistic adversary from that of a truly random function. Formally, a pseudorandom function is defined as:

\begin{definition}[Pseudorandom Function]
A function $f:\{0,1\}^* \times \{0,1\}^* \rightarrow \{0,1\}^*$ is pseudorandom if for all PPT adversaries $\mathcal{A}$, there exists a negligible function $negl$ such that:
\begin{center}
$|Pr[\mathcal{A}^{f_k(\cdot)}=1]-Pr[\mathcal{A}^{F(\cdot)}=1]|<negl(n)$
\end{center}
where $k \rightarrow \{0,1\}^n$ is chosen uniformly randomly and $F$ is a function chosen uniformly randomly from the set of function mapping n-bit strings to n-bit strings.
\end{definition}

Our proof relies on the assumption that the Decisional Diffie-Hellman (DDH) is hard in a group $\mathbb{G}$, i.e., it is hard for an adversary to distinguish between group elements $g^{\alpha \beta}$ and  $g^{\gamma}$ given  $g^{\alpha}$ and  $g^{\beta}$.

\begin{definition}[DDH Assumption]
The DDH problem is hard regarding a group $\mathbb{G}$ if for all PPT adversaries $\mathcal{A}$, there
exists a negligible function $negl$ such that
$|Pr[\mathcal{A}(\mathbb{G},q,g,g^\alpha,g^\beta,g^{\alpha \beta } )=1]-Pr[\mathcal{A}(\mathbb{G},q,g,g^\alpha,g^\beta,g^\gamma)=1]|<negl(k)$
where $\mathbb{G}$ is a cyclic group of order $q$ $(|q| = k)$ and $g$ is a generator of $\mathbb{G}$, and $\alpha, \beta, \gamma \in \mathbb{Z}_q$ are uniformly randomly chosen.
\end{definition}

Encryption algorithms in policy deployment phase are based on \textbf{ClientEnc} and \textbf{ServerReEnc} functions that is equivalent to encrypting a single keyword in the SDE scheme \cite{Dong2011}. Dong \emph{et al.} \cite{Dong2011} show that the single keyword encryption scheme is indistinguishable under chosen plaintext attack (\textit{IND-CPA}). A cryptosystem is considered IND-CPA secure if no PPT adversary, given an encryption of a message randomly chosen from two plaintext messages chosen by the adversary, can identify the message choice with non-negligible probability. Dong \emph{et al.} \cite{Dong2011} prove the following theorem about the single Keyword Encryption (KE) scheme:

\begin{theorem}
If the DDH problem is hard relative to $\mathbb{G}$, then the single keyword encryption
scheme $KE$ is IND-CPA secure against the server $\mathit{S}$, i.e., for all PPT adversaries $\mathcal{A}$ there exists a negligible function $negl$ such that:
\begin{equation}
\begin{array}{l}
Succ_{KE,S}^{\mathcal{A}}(k)=Pr \left[ b'=b \left|
\begin{matrix}
(param, msk) \leftarrow Init(1^k)\\
(K_u,K_s) \leftarrow KeyGen(msk,U)\\
w_0,w_1 \leftarrow \mathcal{A}^{ClientEnc(K_u, \cdot)}(K_s)\\
b \xleftarrow{R} \{0,1\}\\
c_i^*(w_b) = ClientEnc(x_{i1},w_b) \\
b' \leftarrow \mathcal{A}^{ClientEnc(K_u, \cdot)}(K_s,c^*_i(w_b))
\end{matrix}
\right]\right. \\
<\frac{1}{2}+negl(k)
\end{array}
\end{equation}
\end{theorem}

Proof. See Theorem 1 in \cite{Dong2011}.

\subsection{Security of Encryption Algorithms in the Policy Deployment Phase}
Using the fact that the $KE$ scheme is IND-CPA secure, we show that the four encryption schemes: RA, PA, CC and RH are also IND-CPA against the server. We give the proof details for the Roles Assignment encryption scheme $RA$. We will show that the following theorem holds:

\begin{theorem}
If the single keyword encryption $KE$ scheme is IND-CPA secure against the server, then the RA encryption scheme $RA$ is also IND-CPA, i.e., for all PPT adversaries $\mathcal{A}$, there exists a negligible function $negl$ such that
$Succ_{RA,S}^{\mathcal{A}}(k) < \frac{1}{2} + negl(k)$.
\end{theorem}

Proof. We prove the theorem by showing that breaking the $RA$ encryption reduces to breaking the $KE$ encryption. We define the following game in which the adversary $\mathcal{A}$ challenges the game with two lists of roles $L_0$ and $L_1$ having the same number of roles $t$. We construct the following vector containing the encryption of roles from both lists: $\vec{C}^{(i)}=C(r_0^1),\ldots,C(r_0^i),C(r_1^{i+1}),\ldots,C(r_1^t)$. The success probability of the adversary in distinguishing the encryption of the two lists of roles is defined as:

\begin{equation}
Succ_{\mathcal{A}}(k) =\frac{1}{2}
Pr[A(\vec{C}^0) = 0] + \frac{1}{2}
Pr[A(\vec{C}^t) = 1]
\end{equation}

In the following, we show that breaking the $RA$ scheme reduces to breaking the $KE$ game. In the $KE$ game from \cite{Dong2011}, the adversary challenges the game with two keywords $w_0$ and $w_1$ and tries to distinguish between their encryptions. Let us consider a PPT adversary $\mathcal{A}'$ who attempts to challenge the single keyword encryption scheme $KE$ using the corresponding $RA$ adversary $\mathcal{A}$ as a sub-routine The game is the following:
\begin{itemize}
\item $\mathcal{A}'$ is given the parameters $(\mathbb{G},q,g,h,H,f)$ as input and for each user $i$ is given $(i,x_{i2})$.
\item $\mathcal{A}'$ passes these parameters to $\mathcal{A}$.
\item $\mathcal{A}$ generates two lists of roles $L_0$ and $L_1$ having the same number of roles  $t$ and gives them to $\mathcal{A}'$.
\item $\mathcal{A}'$ chooses $i \xleftarrow{r} [1, t]$. It then uses $r^i_0, r^i_1$ to challenge the single keyword encryption $KE$ game. The adversary gets back $c^i_b$ as the result, where $c^i_b$ is the encryption of either $r^i_0$ or $r^i_1$. $\mathcal{A}'$ uses this result to construct a hybrid vector $(c^1_0,\ldots, c^{i-1}_0, c_b^i, c_1^{i+1},\ldots,c^t_1)$ and sends it to $\mathcal{A}$.
\item $\mathcal{A}'$ outputs $b'$, the bit output by $\mathcal{A}$.
\end{itemize}

$\mathcal{A}$ is required to distinguish $\vec{C}^{(i)}$ and $\vec{C}^{(i-1)}$ and the probability of $\mathcal{A}$'s success in distinguishing correctly is:
\begin{equation}
Succ_{\mathcal{A}}^i(k) =\frac{1}{2}
Pr[A(\vec{C}^{(i)}) = 0] + \frac{1}{2}
Pr[A(\vec{C}^{(i-1)}) = 1]
\end{equation}

Because $i$ is randomly chosen, it holds that:
\noindent
\begin{equation}
\begin{array}{lll}
Succ_{\mathcal{A}'}(k) & = &\sum_{i=1}^tSucc_{\mathcal{A}}^i(t) \cdot \frac{1}{t} \\
& = &\frac{1}{2t}Pr[A(\vec{C}^0) = 0] + \sum_{i=1}^{t-1}(Pr[A(\vec{C}^i) = 0] \\
& & +Pr[A(\vec{C}^i) = 1]) + \frac{1}{2}Pr[A(\vec{C}^t) = 1] \\
& = & \frac{1}{t} (\frac{1}{2}Pr[A(\vec{C}^0) = 0] + \frac{1}{2}Pr[A(\vec{C}^t)=1]) + \frac{t-1}{2t} \\
& = & \frac{1}{t}Succ_{\mathcal{A}}(k)+ \frac{t-1}{2t}
\end{array}
\end{equation}

Because the success probability of $\mathcal{A}'$ to break the single keyword encryption scheme is $Succ_{\mathcal{A}'}(k) < \frac{1}{2} + negl(k)$, it follows that $Succ_{\mathcal{A}}(k) < \frac{1}{2} + negl(k)$.

The proof for the other encryption schemes is similar and for lack of space we do not show all the details.

\subsection{Security of Algorithms in the Policy Evaluation Phase}
We now analyse the security of SR, Search Permission, Contextual Condition Evaluation and Search Role Hierarchy. These algorithms require the SP to take some client input (i.e., trapdoors computed using Algorithm \ref{algo:client-td}), process it (i.e., re-encrypt it using Algorithm \ref{algo:server-td}), and test whether it matches some information stored on the server. Though a single operation has been proved secure, we are interested in what these algorithms leak to the SP. We follow the concept of non-adaptive indistinguishability security introduced for encrypted databases by \cite{Curtmola2006} and adapted by \cite{Dong2011} in a multi-user setting. We show that given two non-adaptively generated histories with the same length and outcome, no PPT adversary can distinguish the histories based on what it can observe from the interaction. A history contains all the interactions between clients and the SP. Non-adaptive history means that the adversary cannot choose sequences of client inputs based on previous inputs and matching outcomes.

In the following, we show the details for the SR scheme. In this scheme, a history is defined as follows:

\begin{definition}[SR History]
An SR history $\mathcal{H}_i$ is an interaction between a SP and all clients that connect to it, over $i$ role activation requests. $\mathcal{H}_i=(L_s^{u_1},\ldots, L_s^{u_i},r_{1}^{u_1},\ldots,r_{i}^{u_i})$, where $u_i$ represents an identifier of the client making the requests, $L_s^{u_i}$ represents the lists of roles for client $u_i$, and $r_{i}^{u_i}$ represents the request made by the client.
\end{definition}

We formalise the information leaked to a SP as a \textit{trace}. We define two kinds of traces: the trace of a single request and the trace of a history. The trace of a request leaks to the SP which role in $L_s^{i}$ matches the request and can be formally defined as: $tr(r)=\{td*_i(role), L_s^{i}, idx\}$, where $idx$ is the index of the matched role, if any, in $L_s^{i}$.

We define the role matching pattern $\mathcal{P}$ over a history $\mathcal{H}_i$ to be a set of  binary matrices (one for each client) with columns corresponding to encrypted roles in the list of the client, and rows corresponding to requests. $\mathcal{P}[j,k]=1$ if request $j$ matched the $k$'s role and $\mathcal{P}[j,k]=0$ otherwise.

The trace of a history includes the encrypted role assignment lists of all clients $L_s^{u_i}$ stored by the SP and which can change as new roles are added and clients leave of join the system, the trace of each request, and the role matching pattern $\mathcal{P}_i$ for each client.

During an interaction, the adversary cannot see directly the plaintext of the request, instead it sees the ciphertext. The view of a request is defined as:

\begin{definition}[View of a Request]
We define the view of a request  $q_{1}^{u_1}$ under a key set $K_{ui}$ as: 
$V_{K_{ui}}(q^{u_i})= tr(q^{u_i})$
\end{definition}

\begin{definition}[View of a History]
We define the view of a history with $i$ interactions $\mathcal{H}_i$ as $V_{K_u}(H_i)=(L_s^{u_1},\ldots,L_s^{u_i},V_{K_{ui}}(q_1^{u_i}),\ldots,V_{K_{ui}}(q_i^{u_i})$.
\end{definition}

The security definition is based on the idea that the scheme is secure if nothing is leaked to the adversary beyond what the adversary can learn from traces.

We define the following game in which an adversary $\mathcal{A}$ generates two histories $\mathcal{H}_{i0}$ and $\mathcal{H}_{i1}$ with the same trace over $i$ requests. Then the adversary is challenged to distinguish the views of the two histories. If the adversary succeeds with negligible probability, the scheme is secure.

\begin{definition}[Non-adaptive indistinguishability against a curious SP]
The SR scheme is secure in the sense of non-adaptive indistinguishability against a curious SP if for all $i \in \mathbb{N}$ and for all PPT adversaries $\mathcal{A}$ there exists a negligible function $negl$ such that:
\begin{equation}
Pr \left[ b'=b \left|
\begin{array}{lll}
(params, msk) \leftarrow Init(1^k)\\
(K_u,K_s) \leftarrow KeyGen(msk,U)\\
\mathcal{H}_{io},\mathcal{H}_{i1} \leftarrow \mathcal{A}(K_s)\\
b \xleftarrow{R} \{0,1\}\\
b' \leftarrow \mathcal{A}(K_s,V_{K_u}(\mathcal{H}_{ib}))
\end{array}
\right]<\frac{1}{2}+negl(k)
\right.
\end{equation}
where U is a set of user IDs, $K_u$ is the user side key sets, $K_s$ are the server side key sets, $\mathcal{H}_{i1}$ and $\mathcal{H}_{i0}$ are two histories over $i$ requests such that $Tr(\mathcal{H}_{i0}) = Tr(\mathcal{H}_{i1})$.
\end{definition}

\begin{theorem}\label{thm:3}
If the DDH problem in hard relative to $\mathbb{G}$, then the SR scheme is a non-adaptive indistinguishable secure scheme. The success probability of a PPT adversary $\mathcal{A}$ in breaking the SR scheme is defined as:
\begin{equation}
\begin{array}{l}
Succ^{\mathcal{A}}(k)=  \frac{1}{2}Pr[\mathcal{A}(RA(\vec{L}_0),TD(\vec{r}_0))=0]+ \\
\hspace{50pt} \frac{1}{2}Pr[\mathcal{A}(RA(\vec{L}_1),TD(\vec{r}_1))=1] \\
\hspace{41pt} < \frac{1}{2} + negl(k)
\end{array}
\end{equation}
where $RA(\vec{L}_i)$ is the role encryption of the vector of lists of $H_i$, and $TD(\vec{r}_i)$ is the \textbf{ClientTD} of the roles in the requests of $H_i$.
\end{theorem}

Proof. We consider an adversary $\mathcal{A}'$ that challenges the RE IND-CPA game using $\mathcal{A}$ as a sub-routine. $\mathcal{A}'$ does the following:
\begin{itemize}
\item $\mathcal{A}'$ receives public parameters $params$ and the server side $(i,x_{i2})$ keys.

\item To generate a view of a history $\mathcal{H}_i=(L_1^{u_1},\ldots,L_i^{u_i}, q_1^{u_1},\ldots, q_i^{u_i})$. $\mathcal{A}'$ performs the following steps:
\begin{itemize}
\item For each role assignment list $L_j^{u_j}$, run Algorithm 5 to encrypt it as $RA(L_j^{u_j})$. 
\item For each Search Role request $q_j^{u_j}$, run $ClientTD$ to generate the trapdoor $TD(r)$ for the role.
\end{itemize}

\item $\mathcal{A}$ outputs $\mathcal{H}_{i0},\mathcal{H}_{i1}$. $\mathcal{A}'$ encrypts $\mathcal{H}_{i1}$ by itself and challenges the RE IND-CPA game with $\vec{L}_0$ and $\vec{L}_1$, the vectors of all roles lists in the two histories. It gets the result $RA(\vec{L}_b)$ where $b \xleftarrow{R} \{0,1\}$ and forms a view of a history $(RA(\vec{L}_b), TD(\vec{r_1}))$. It sends the view to $\mathcal{A}$.

\item $\mathcal{A}$ tries to determine which vector was encrypted and outputs $b' \in \{0,1\}$.

\item $\mathcal{A'}$ outputs $b'$.

\end{itemize}

Because the $RA$ scheme is IND-CPA, it follows that:

\begin{equation}
\begin{array}{l}
\frac{1}{2} + negl(k)  >  Succ_{RA}^{\mathcal{A}'}(k) \\
\hspace{47pt} = \frac{1}{2}Pr[\mathcal{A}((RA(\vec{L}_0),TD(\vec{r}_1)))=0]+\\
\hspace{56pt} \frac{1}{2}Pr[\mathcal{A}((RA(\vec{L}_1),TD(\vec{r}_1)))=1]
\end{array}
\end{equation}

Now let us consider another adversary $\mathcal{A}''$ who wants to distinguish the
pseudorandom function $f$ using $\mathcal{A}$ as a sub-routine. The adversary does the following:
\begin{itemize}
\item It generates $(\mathbb{G}, q, g, h, H)$ as public parameters, and sends them to $\mathcal{A}$ along with $f$. For each user $i$, it chooses randomly $x_{i1}$, $x_{i2}$ such that $x_{i1} + x_{i2} = x$. It sends all $(i, x_{i2})$ to $\mathcal{A}$ and keeps all $(i, x_{i1}, x_{i2})$.
\item $\mathcal{A}$ outputs $\mathcal{H}_{i0},\mathcal{H}_{i1}$. $\mathcal{A}''$ encrypts all the roles lists in $\mathcal{H}_{i0}$ as $RA(\vec{L}_0)$. It chooses $b \xleftarrow{R} \{0,1\}$ and asks the oracle to encrypt all roles in $\mathcal{H}_{ib}$. It combines the results to form a view $(RA(\vec{L}_0),TD(\vec{r}_b))$ and returns it to $\mathcal{A}$.
\item $\mathcal{A}$ outputs $b'$. $\mathcal{A}''$ outputs $1$ if $b'=b$ and $0$ otherwise.
\end{itemize}

There are two cases to consider:
Case 1: the oracle in $\mathcal{A}''$s game is the pseudorandom function $f$, then:

\begin{equation}
\begin{array}{l}
Pr[\mathcal{A}''^{f_{s}(.)}(1^k)=1] = \\
\hspace{50pt} \frac{1}{2}Pr[\mathcal{A}(RA(\vec{L}_0),TD(\vec{r}_0))=0]+\\
\hspace{50pt} \frac{1}{2}Pr[\mathcal{A}(RA(\vec{L}_0),TD(\vec{r}_1))=1]
\end{array}
\end{equation}

Case 2: the oracle in $\mathcal{A}''$s game is a random function $F$, then for each distinct
role $r$, $\sigma_r$ is completely random to $\mathcal{A}$. Moreover, we know the traces are identical, so $RA(\vec{L}_b)$ and $TD(\vec{r}_b)$ are completely random to $\mathcal{A}$. In this case:
\begin{equation}
Pr[\mathcal{A}''^{f_{s}(.)}(1^k)=1]=\frac{1}{2}
\end{equation}

Because $f$ is a pseudorandom function, by definition it holds that:
\begin{equation}
\begin{array}{l}
|Pr[\mathcal{A}''^{f_{s}(.)}(1^k)=1]-Pr[\mathcal{A}'^{f_{s}(.)}(1^k)=1]|<negl(k)\\
\hspace{88pt}Pr[\mathcal{A}''^{f_{s}(.)}(1^k)=1]<\frac{1}{2}+negl(k)
\end{array}
\end{equation}

Sum up $Succ_{RE}^{\mathcal{A}'}(k)$ and $Pr[\mathcal{A}''^{f_{s}(.)}(1^k)=1]$:

\begin{equation}
\begin{array}{l}
1+negl(k) >  \frac{1}{2}Pr[\mathcal{A}(RA(\vec{L}_0),TD(\vec{r}_0))=0]+\\
\hspace{56pt}\frac{1}{2}Pr[\mathcal{A}(RA(\vec{L}_0),TD(\vec{r}_1))=1]+\\
\hspace{56pt}\frac{1}{2}Pr[\mathcal{A}(RA(\vec{L}_0),TD(\vec{r}_1))=0]+ \\
\hspace{56pt}\frac{1}{2}Pr[\mathcal{A}(RA(\vec{L}_1),TD(\vec{r}_1))=1] \\
\hspace{47pt}=\frac{1}{2}Pr[\mathcal{A}(RA(\vec{L}_0),TD(\vec{r}_0))=0]+\\
\hspace{56pt}\frac{1}{2}+\\
\hspace{56pt}\frac{1}{2}Pr[\mathcal{A}(RA(\vec{L}_1),TD(\vec{r}_1))=1]+ \\
\hspace{47pt}= \frac{1}{2} + Succ^{\mathcal{A}}(k)
\end{array}
\end{equation}

Therefore $Succ^{\mathcal{A}}(k)< \frac{1}{2} + negl(k)$.

%\section{Discussion}
%\label{sec:discussion}
%This section provides the discussion about security aspects of $\mathit{ESPOON_{ERBAC}}$.

\subsection{Revealing Policy Structure}
The policy structure reveals information about the operators, such as AND and OR, and the number of operands used in the contextual condition. To overcome this problem, dummy attributes could be inserted in the tree representing contextual conditions. Similarly, the PIP can send dummy attributes to the PDP at the time of policy evaluation to obfuscate the number of attributes required for evaluating any contextual condition.

\section{Performance Analysis}
\label{sec:performance-analysis}

In this section, we discuss a quantitative analysis of the performance of $\mathit{ESPOON_{ERBAC}}$. It should be noticed that here we are concerned about quantifying the overhead introduced by the encryption operations performed both at the trusted environment and the outsourced environment. In the following discussion, we do not take into account the latency introduced by the network communication.

\subsection{Implementation Details}

We have implemented $\mathit{ESPOON_{ERBAC}}$ in Java $1.6$. We have developed all the components of the architecture required for performing the policy deployment and policy evaluation phases. For the cryptographic operations, we have implemented all the functions presented in Section \ref{sec:algorithmic-details}. We have tested the implementation of $\mathit{ESPOON_{ERBAC}}$ on a single node based on an Intel Core2 Duo $2.2$ GHz processor with $2$ GB of RAM, running Microsoft Windows XP Professional version $2002$ Service Pack $3$. %The number of iterations performed for each of the following results is $1000$.

\begin{figure*}
\centering
\subfigure[]{
% left bottom right top
\includegraphics[trim=15mm 15mm 135mm 220mm,clip,width=.31\textwidth]{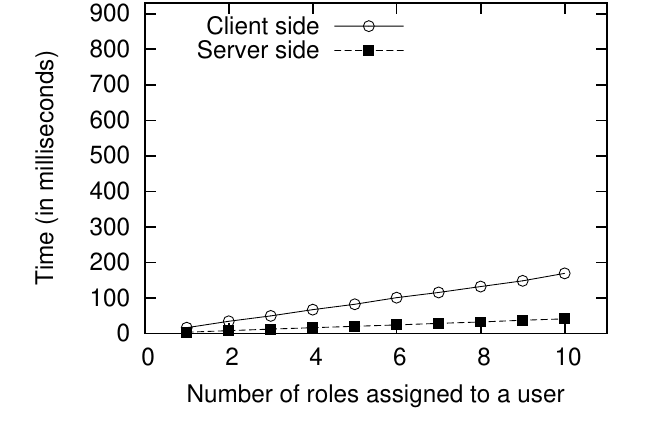} % .3
\label{fig:deploy-ura}
}
\subfigure[]{
\includegraphics[trim=15mm 15mm 135mm 220mm,clip,width=.31\textwidth]{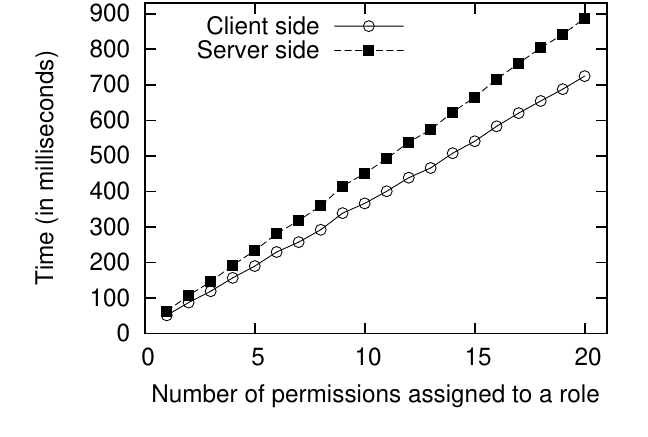} % .3
\label{fig:deploy-rpa}
}
\subfigure[]{
\includegraphics[trim=15mm 15mm 135mm 220mm,clip,width=.31\textwidth]{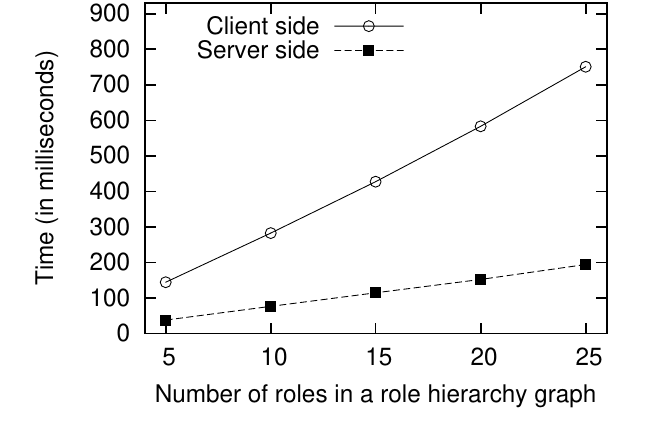} % .3
\label{fig:deploy-role-hierarchy}
}
\caption{Performance overhead of deploying RBAC policies: \subref{fig:deploy-ura} roles assigned to a user, \subref{fig:deploy-rpa} permissions to a role and \subref{fig:deploy-role-hierarchy} a role hierarchy graph}
%\caption[Performance overhead of contextual conditions during the policy deployment]%{Caption of subfigures \subref{fig:subfig1}, \subref{fig:subfig2} and \subref{fig:subfig3}}
\label{fig:policy-deployment-rbac-policy}
\end{figure*}

\subsection{Performance Analysis of the Policy Deployment Phase}
In this section, we analyse the performance of the policy deployment phase. In this phase, an Admin User encrypts policies and sends those encrypted policies to the Administration Point running in the outsourced environment. The Administration Point re-encrypts policies and stores them in the Policy Store in the outsourced environment. In the following, we analyse the performance of deploying (part of) policies including role assignment, permission assignment, contextual conditions and role hierarchy graph.

\emph{\textbf{Role Assignment:}} 
In order to deploy a role assignment policy, an Admin User performs a first round of encryption on the client side (see Algorithm \ref{algo:deploy-role-assignment-client-side}) and sends the client encrypted role assignment policy to the Administration Point. The Administration Point performs another round of encryption on the server side (see Algorithm \ref{algo:deploy-role-assignment-server-side}) before storing role assignment policy in the Policy Store. Figure \ref{fig:deploy-ura} shows performance overhead on the client side, as well as on the server side in order to deploy a role assignment policy. In this graph, we observe the performance by increasing number of roles in a role assignment policy. As we can expect, the performance overhead increases linearly with the linear increase in the number of roles in a role assignment policy. As we can notice, the graph grows linearly with the linear increase in the number of roles in the role assignment policy.

During the policy deployment phase, the encryption algorithm on the client side (Algorithm \ref{algo:client-enc}) takes more time that of the server side (Algorithm \ref{algo:server-re-enc}) as shown in Figure \ref{fig:policy-deployment-rbac-policy}. The encryption algorithm on the client side takes more time because it performs more complex cryptographic operations such as random number generation and hash calculation as illustrated in Algorithm \ref{algo:client-enc}. However, any policy is deployed very rarely; whereas, it may be evaluated quite frequently. Therefore, the performance overhead of the policy evaluation phase (discussed in Section \ref{sec:policy-evaluation}) is of great importance.

\emph{\textbf{Permission Assignment:}} 
For deploying permissions to a role, an Admin User performs a first round of encryption on the client side (see Algorithm \ref{algo:deploy-permission-assignment-client-side}) and sends both the client encrypted role and client encrypted permissions to the Administration Point, where each permission contains both an action and a target. The Administration Point generates the server encrypted role and server encrypted permissions after performing a second round of encryption on the server side (see Algorithm \ref{algo:deploy-permission-assignment-server-side}). Figure \ref{fig:deploy-rpa} shows the performance overhead of deploying a permission assignment policy. This graph illustrates the performance of deploying a permission assignment policy for a role with a number of permissions ranging from 1 to 20. As we can expect, the performance overhead increases linearly with the linear increase in the number of permissions in the permission assignment policy.

\emph{\textbf{Contextual Conditions:}} 
Both role assignment and permission assignment policies include a contextual condition as we can see in Figure \ref{fig:policy-role-assignment} and Figure \ref{fig:policy-permission-assignment}, respectively. The contextual condition is represented as a tree structure as illustrated in Figure \ref{fig:cc}. During the policy deployment phase, an Admin User encrypts each leaf node of the tree (see Algorithm \ref{algo:deploy-contextual-condition-client-side}) while the Administration Point re-encrypts each leaf node (see Algorithm \ref{algo:deploy-contextual-condition-server-side}) and finally stores the tree in the Policy Store either in the Role Repository or the Permission Repository. 

%TODO: adjust $ \\ $\mathit on paragraph change
In the tree representing contextual conditions, leaf nodes represent string comparisons (for instance, $\mathit{Location = Cardiology \mhyphen ward}$) and/or numerical comparisons (for instance, $\mathit{Access Time}$ \\ $\mathit{> 9}$). A string comparison is always represented by a single leaf node while a numerical comparison may require more than one leaf nodes. In the worst case, a single numerical comparison, represented as $s$ bits, may require $s$ separate leaf nodes. Therefore, numerical comparisons have a major impact on the encryption of a policy at deployment time.

\begin{figure*}
\centering
\subfigure[]{
\includegraphics[trim=15mm 15mm 125mm 210mm,clip,width=.48\textwidth]{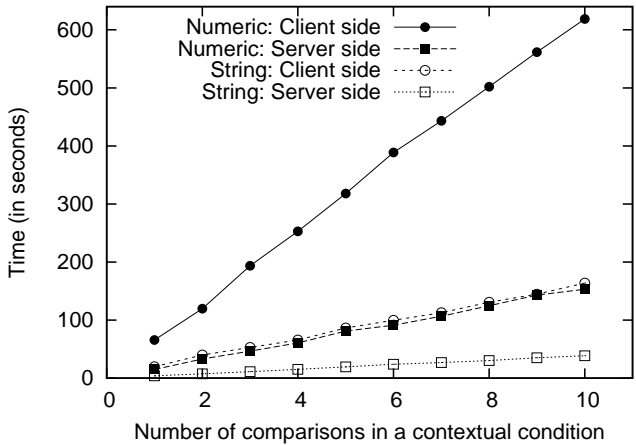} % .35
\label{fig:deploy-context-attr}
}
\subfigure[]{
\includegraphics[trim=15mm 15mm 125mm 210mm,clip,width=.48\textwidth]{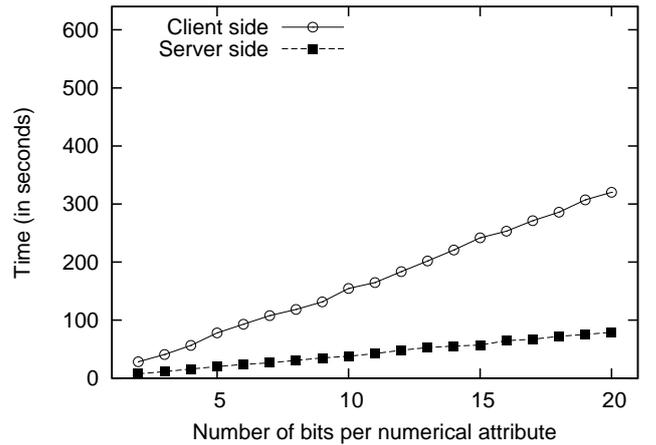} % .35
\label{fig:deploy-context-attr-bits}
}
\caption{Performance overhead of deploying contextual conditions: \subref{fig:deploy-context-attr} numerical and string comparisons and \subref{fig:deploy-context-attr-bits} size of a numerical attribute}
%\caption[Performance overhead of contextual conditions during the policy deployment]%{Caption of subfigures \subref{fig:subfig1}, \subref{fig:subfig2} and \subref{fig:subfig3}}
\label{fig:policy-deployment-context}
\end{figure*}

Figure \ref{fig:deploy-context-attr} illustrates the performance overhead of deploying numerical and string comparisons. In this graph, we increase the number of string comparisons and numerical comparisons present in the contextual condition of a policy. As the graph, the time taken by deployment functions on the client side and the server side grow linearly with the number of comparisons in the contextual condition. The numerical comparisons have a stepper line because one numerical comparison of size $s$ may be equivalent to $s$ string comparisons in the worst case. For string comparisons, we have used ``$attributeName_i$=$attributeValue_i$'', where $i$ varies from 1 to 10. For numerical comparisons, we have used ``$attributeName_i<15$\#4''.\footnote{It should be noted that using the comparison less than 15 in a 4-bit representation represents the worst case scenario requiring 4 leaf nodes.}

To check how the size of the bit representation impacts on the encryption functions during the deployment phase, we have performed the following experiment. We fixed the number of numerical comparisons in the contextaul condition to only one and increased the size $s$ of the bit representation from $2$ to $20$ for the comparison ``$attributeName<2^s-1$. Figure \ref{fig:deploy-context-attr-bits} shows the performance overhead of the encryption during the policy deployment phase on the client side, as well as on the server side. We can see that the policy deployment time incurred grows linearly with the increase in the size $s$ of a numerical attribute. In general, the time complexity of the encryption of the contextual conditions during the policy deployment phase is $O(m+ns)$ where $m$ is the number of string comparisons, $n$ is the number of numerical comparisons, and $s$ represents the number of bits in each numerical comparison.

\emph{\textbf{Role Hierarchy Graph:}} 
The PDP may search for a base role of the one in the access request $\mathit{REQ}$ since a derived role inherits all permissions from its base role. For supporting this search, we deploy a role hierarchy graph. For deploying a role hierarchy graph, an Admin User performs the first round in order to generate the client encrypted trapdoor, as well as to calculate the client generated trapdoor of each role in the graph (see Algorithm \ref{algo:deploy-role-hierarchy-client-side}). The Admin User sends the client generated role hierarchy graph to the Administration Point. The Administration Point performs the second round to generate the server encrypted trapdoor, as well as to calculate the server generated trapdoor of each role in the graph (see Algorithm \ref{algo:deploy-role-hierarchy-server-side}). The PDP matches the trapdoor of role in $\mathit{REQ}$ with the server encrypted role and if this match is successful, it finds trapdoors of the base roles. The trapdoors of base roles are required in order to perform search in the list of server encrypted roles in the Permission Repository.

In our experiment, we consider a role hierarchy graph in which each role $R_i$ extends role $R_{i+1}$ for all values of $i$ from 0 to $n - 1$ where $n$ indicates the total number of nodes and varies from 5 to 25. Figure \ref{fig:deploy-role-hierarchy} shows the performance overhead of encrypting a role hierarchy graph both on the client side and the server side. The graph grows linearly with the number of roles in a role hierarchy graph.

\begin{table}[htp]
\centering
\caption{Performance overhead of encrypting requests during the policy evaluation phase}
\label{tab:request}
{\small
\begin{tabular}{ |c|c| }
\hline
\textbf{Request Type} & \textbf{Time (in milliseconds)} \\ \hline
$\mathit{ACT}$ & 16.353 \\ \hline
$\mathit{REQ}$ & 47.069 \\ \hline
\end{tabular}
}
\end{table}

\subsection{Performance Analysis of the Policy Evaluation Phase}
\label{sec:policy-evaluation}

In this section, we analyse the performance of the policy evaluation phase. In this phase, a Requester sends the encrypted request to the PEP running in the outsourced environment. The PEP forwards the encrypted request to the PDP. The PDP has to select the set of policies that are applicable to the request. The PDP may require contextual information in order to evaluate the selected policies. In the following, we calculate the performance overhead of generating requests, search a role (in the Role Repository, in the Active Roles repository or in the Permission Repository), searching a permission, evaluating contextual conditions and searching a role in a role hierarchy graph.

% access request
\begin{figure*}
\centering
\subfigure[]{
%\includegraphics[width=.235\textwidth]{graphs/request-check-req-role-in-session}
%\label{fig:request-check-req-role-in-session}
% left bottom right top
\includegraphics[trim=15mm 15mm 145mm 225mm,clip,width=.4\textwidth]{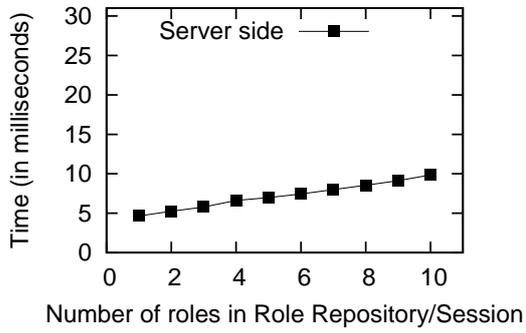} % .235
\label{fig:search-role}
}
\subfigure[]{
\includegraphics[trim=15mm 15mm 145mm 225mm,clip,width=.4\textwidth]{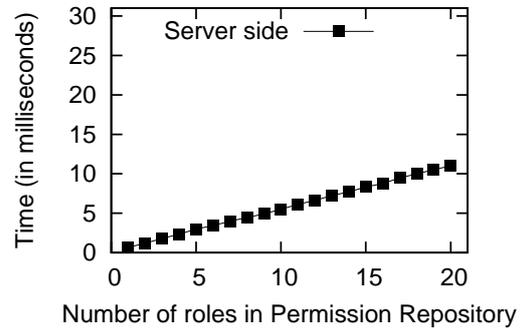} % .235
\label{fig:request-search-req-role-in-perms}
}
\subfigure[]{
\includegraphics[trim=15mm 15mm 145mm 225mm,clip,width=.4\textwidth]{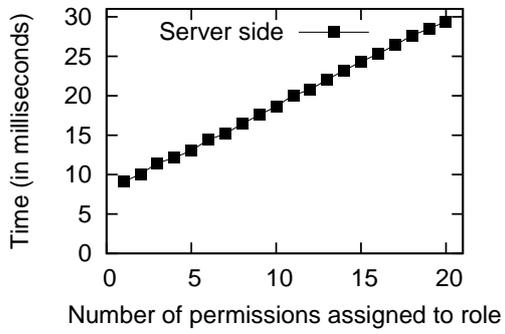} % .235
\label{fig:request-search-req-perms}
}
\subfigure[]{
\includegraphics[trim=15mm 15mm 145mm 225mm,clip,width=.4\textwidth]{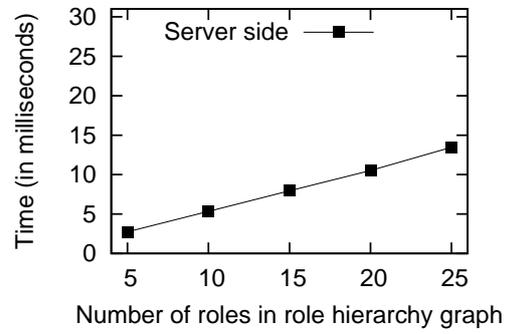} % .235
\label{fig:search-role-hierarchy}
}
\caption{Performance overhead of evaluating RBAC policies}
%\caption[Performance overhead of contextual conditions during the policy deployment]%{Caption of subfigures \subref{fig:subfig1}, \subref{fig:subfig2} and \subref{fig:subfig3}}
\label{fig:policy-evaluation-rbac-policy}
\end{figure*}

% context evaluation
\begin{figure*}
\centering
\subfigure[]{
% left bottom right top
\includegraphics[trim=15mm 15mm 130mm 215mm,clip,width=.48\textwidth]{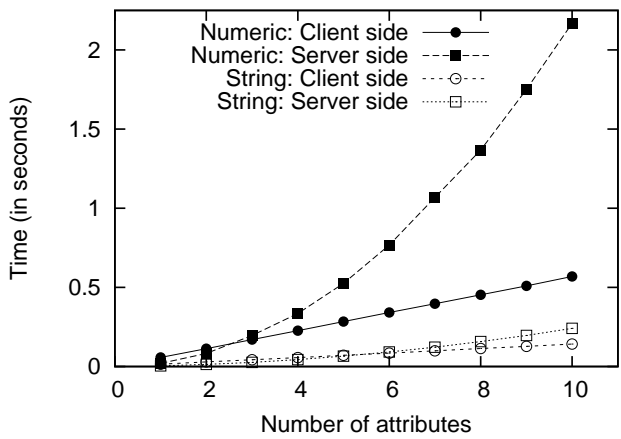} % .35
\label{fig:request-context-attr}
}
\subfigure[]{
\includegraphics[trim=15mm 15mm 130mm 215mm,clip,width=.48\textwidth]{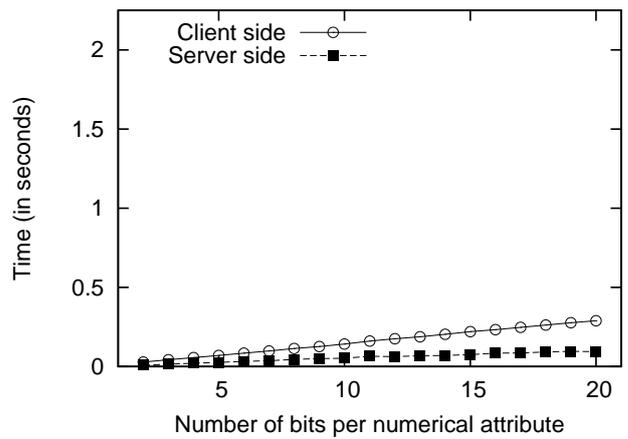} % .35
\label{fig:request-context-attr-bits}
}
\caption{Performance overhead of evaluating contextual conditions}
%\caption[Performance overhead of contextual conditions during the policy deployment]%{Caption of subfigures \subref{fig:subfig1}, \subref{fig:subfig2} and \subref{fig:subfig3}}
\label{fig:policy-evaluation-context}
\end{figure*}

\emph{\textbf{Generating Requests:}}
A Requester may send the role activation request $\mathit{ACT}$. In order to generate $\mathit{ACT}$, a Requester calculates the client generated role (see Algorithm \ref{algo:client-td}). This trapdoor generation of role takes 16.353 milliseconds as illustrated in Table \ref{tab:request}. After a Requester is active in a role, she may make an access request $\mathit{REQ}$ . A Requester has to calculate trapdoor for each element (including role, action and target) in $\mathit{REQ}$. The $\mathit{REQ}$ generation takes 47.069 milliseconds as illustrated in Table \ref{tab:request}. We can see that $\mathit{REQ}$ generation takes 3 times of $\mathit{ACT}$ generation because $\mathit{REQ}$ has to calculate 3 trapdoors while $\mathit{ACT}$ has to generate only a single trapdoor. The request generation does not depend on any parameters and can be considered constant.

\emph{\textbf{Searching a Role in Role Repository/Session:}}
In order to grant $\mathit{ACT}$, the PDP needs to search roles in the Role Repository. For searching a role, the PDP first calculates the server generated trapdoor of role in $\mathit{ACT}$ and then matches this server encrypted trapdoor with server encrypted roles in the role assignment list as illustrated in Algorithm \ref{algo:search-role}. Figure \ref{fig:search-role} shows the performance overhead (in the worst case) of performing this search. In this graph, we can observe that it grows linearly with increase in number of roles. As the graph indicates, the search function takes initial approximately $4$ milliseconds to generate the server encrypted trapdoor of role in $\mathit{ACT}$ while it takes approximately $0.6$ milliseconds to perform encrypted match.

The PDP grants $\mathit{ACT}$ by adding the server encrypted role of the Requester in the Active Roles repository of the Session. This implies that the Session maintains a list of active roles. Once a Requester makes an access request $\mathit{REQ}$, the PDP has to search in the Session if she is already active in role indicated in $\mathit{REQ}$. The performance overhead of searching a role in session is same as it incurs for searching a role in the Role Repository (shown in Figure \ref{fig:search-role}).

\emph{\textbf{Searching a Role in Permission Repository:}}
After finding the role of $\mathit{REQ}$ in the list of active roles, the PDP has to search if the same role has the requested permission. For this purpose, the PDP has first to search the role of $\mathit{REQ}$ in the Permission Repository and if any match is found, it has to search the requested permission in the list of permissions assigned to the found role. Figure \ref{fig:request-search-req-role-in-perms} shows the performance overhead (in the worst case) of searching a role in the Permission Repository. The graph grows linearly with the increase in the number of roles in the Permission Repository. The PDP runs Algorithm \ref{algo:search-role} but with a slight modification of ignoring the server trapdoor generation (in Line \ref{line:search-role-td}) as it is already generated when the role of $\mathit{REQ}$ is searched in the session. This is why, searching a role in the Permission Repository (as illustrated in Figure \ref{fig:request-search-req-role-in-perms}) takes less time than searching a role in the Role Repository or Session (as illustrated in Figure \ref{fig:search-role}).

\emph{\textbf{Searching a Permission:}}
After a role is found in the Permission Repository, the PDP searches the requested permission in the list of permissions assigned to the found role (see Algorithm \ref{algo:search-permission}). Before searching the list of permissions, the PDP has to calculate server generated trapdoors of both the action and the target present in $\mathit{REQ}$. As we explained earlier, a single trapdoor generation on the server side takes approximately 4 milliseconds. The trapdoor generation of the requested permission, containing an action and a target, takes 8 milliseconds. Next, the PDP match (server generated trapdoors of) this requested permission with the list of (sever encrypted) permissions assigned to the found role. Figure \ref{fig:request-search-req-perms} shows the performance overhead (in the worst case) of searching server generated trapdoor of permission with a list of server encrypted permissions. The graph grows linearly with the increase in the number of permissions in the list. For each permission match, the PDP performs (at most) two encrypted matches each incurring approximately 0.6 milliseconds.

%\emph{\textbf{Generating and Evaluating Contextual Attributes:}}

\emph{\textbf{Evaluating Contextual Conditions:}}
For evaluating role assignment (illustrated in Figure \ref{fig:policy-role-assignment}) or permission assignment (illustrated in Figure \ref{fig:policy-permission-assignment}) policies, the PDP may need to evaluate contextual conditions. For evaluating contextual conditions, the PDP needs to fetch contextual information from the PIP. The The PIP is responsible to collect and send the required contextual information that include information about the Requester (for instance, Requester's location or Requester's age) or the environment in which the request is made (for instance, time or temperature). The PIP transforms these attributes into trapdoors before sending to the PDP (as illustrated in Algorithm \ref{algo:request-contextual-condition}). For each single string attribute (for instance, $\mathit{Location:=Cardiology \mhyphen ward}$), the PIP generates a single trapdoor. For each numerical attribute of size s-bit (for instance, $\mathit{Access Time =: 10\#5}$), the PIP generates s trapdoors. Figure \ref{fig:request-context-attr} shows the performance overhead of generating trapdoors by the PIP on the client side for both numerical and string attributes. In our experiment, we vary number of attributes (both string and numeric) from 1 to 10. As we can see, the graph grows linearly with the increase in number of attributes. For numerical attributes, the curve of trapdoor generation on the client side is steeper than that of the string attributes because numerical attribute is of size s bits where s is set to 4. This means that each numerical attribute requires 4 trapdoors; on the other hand, a string attribute requires only a single attribute. We observe also the behaviour of generating client trapdoors for a numerical attribute of varying size. Figure \ref{fig:request-context-attr-bits} shows behaviour of generating on the client side trapdoors of a numerical attribute of varying size ranging from 2 to 20 bits. This graph grows linearly with the increase in number of bits, representing size of a numerical attribute.

After receiving trapdoors of contextual information, the PDP may evaluate a contextual condition. To evaluate the tree representing a contextual condition, the PDP matches contextual information against the leaf nodes in the tree, as illustrated in Algorithm \ref{algo:match-contextual-condition}. To quantify the performance overhead of this encrypted matching, we have performed the following test. First, we have considered two cases: the first case is the one in which the PIP provides only string attributes and the contextual condition contains only string comparisons; in the second, the PIP provides only numerical attributes and the contextual condition consists only of numerical comparisons. For both cases, the number of attributes varies together with the number of comparisons in the tree. In particular, if the PIP provides $n$ different attributes then the contextual condition will contain $n$ different comparisons.

Figure \ref{fig:request-context-attr} shows also the performance overhead of evaluating string and numerical comparisons on the server side. As we can see, the condition evaluation for numerical attributes has a steeper curve. This can be explained as follows. For the first case, for each string attribute only a single trapdoor is generated. A string comparison is represented as a single leaf node in the tree representing a contextual condition. This means that $n$ trapdoors in a request are matched against $m$ leaf nodes in the tree resulting in a $O(nm)$ complexity (however, in our experiments the number of attributes and the number of comparisons are always the same). For the case of the numerical attributes, we have also to take in to consideration the bit representation. In particular, for a give numerical attribute represented as $s$ bits, we need to generate $s$ different trapdoors. This means that $n$ numerical attributes in a request will be converted in to $n s$ different trapdoors. These trapdoors then need to be matched against the leaf nodes representing the numerical comparisons. Figure \ref{fig:request-context-attr-bits} shows the performance overhead of evaluating a numerical comparison where the size of a numerical attribute varies from 2 to 20. As we have discussed for the policy deployment phase, in the worst case scenario, a numerical comparison for a $s$-bit numerical attribute requires $s$ different leaf nodes. In a tree with $m$ different numerical comparisons, this means that the $n s$ trapdoors need to be matched against $m s$ resulting in $O(n m s^2)$ complexity.

\emph{\textbf{Searching a Role Hierarchy Graph:}}
The PDP may search a role in the role hierarchy graph. For performing this search, we consider a role hierarchy graph in which each role $R_i$ extends role $R_{i+1}$ for all values of $i$ from 0 to $n - 1$ where $n$ indicates the total number of nodes and varies from 5 to 25. Figure \ref{fig:search-role-hierarchy} shows the performance overhead of searching a role in the role hierarchy graph deployed on the server side. As we can expect, the graph grows linearly with the number of roles in a role hierarchy graph.

\emph{\textbf{Comparing $\mathit{ESPOON_{ERBAC}}$ with $\mathit{ESPOON}$:}}
We compare the performance overheads of the policy evaluation of $\mathit{ESPOON_{ERBAC}}$ with that of $\mathit{ESPOON}$ \cite{Asghar2011ARES}. Before we show the comparison, we see how policies are expressed in both $\mathit{ESPOON_{ERBAC}}$ and $\mathit{ESPOON}$. The $\mathit{ESPOON_{ERBAC}}$ policies are explained in Section \ref{sec:representation}. The $\mathit{ESPOON}$ policy is expressed as a $\langle S, A, T \rangle$ tuple with a $\mathit{CONDITION}$, meaning if $\mathit{CONDITION}$ holds then subject $S$ can take action $A$ over target $T$. For comparing the performance overheads, we consider $\mathit{ESPOON}$ policies with 50 unique subjects and each subject has 10 unique actions and targets where each $\langle S, A, T \rangle$ tuple's condition is the conjunction (AND) of the contextual condition illustrated in Figure \ref{fig:cc} and \emph{RequesterName$=<$NAME$>$}. That is, a subject can execute action over the target provided subject's name is equal to one specified in the condition, subject's location is cardiology-ward and time is between 9 AM and 5 PM. Similarly, we consider $\mathit{ESPOON_{ERBAC}}$ policies with 50 unique roles and each role has 10 unique permissions, where each user can get active in 5 roles. The introduction of RBAC simplifies the roles and permission management because we can enforce possible conditions at role activation time instead of enforcing them at the permission grant time. For instance, we can enforce location and time checks (i.e., the condition illustrated in Figure \ref{fig:cc}) at the role activation time while the condition \emph{RequesterName$=<$NAME$>$} can be enforced at the permission grant time. 

\begin{figure}
\centering
% left bottom right top
\includegraphics[trim=15mm 10mm 90mm 185mm,clip,width=.45\textwidth]{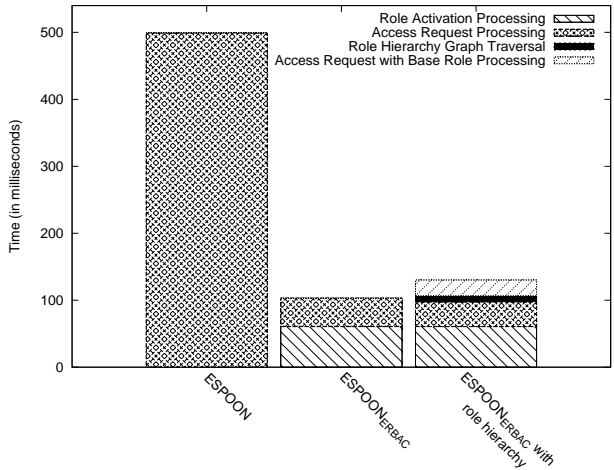} % .3
\caption{Performance comparison of $\mathit{ESPOON}$ and $\mathit{ESPOON_{ERBAC}}$}
\label{fig:espoon-vs-erbac}
\end{figure}

Figure \ref{fig:espoon-vs-erbac} shows the performance overheads of evaluating $\mathit{ESPOON}$ and $\mathit{ESPOON_{ERBAC}}$ policies. In $\mathit{ESPOON}$, a requester's subject is matched with one in the repository of 500 entries (i.e., 50 subjects each with 10 actions and targets). If there is any match, requester's action and target are matched and then condition is evaluated. In the worst case, in $\mathit{ESPOON}$, the access request processing can take approximately up to 500 milliseconds. On the other hand, in $\mathit{ESPOON_{ERBAC}}$, a requester first gets active in a role provided condition holds. The role activation can take approximately up to 60 milliseconds for a user that can get active in 5 roles. After the role activation, a requester can be granted permissions assigned to its role. However, first the active role is searched in the session and then the permission can be granted if the condition associated with that permission holds. As we can see in Figure \ref{fig:espoon-vs-erbac}, grating the permission takes up to 42 milliseconds. The reason why $\mathit{ESPOON_{ERBAC}}$ performance is better than that of $\mathit{ESPOON}$ because (i) all possible conditions are enforced at the role activation time and (ii) introduction of roles simplified the roles and permissions management.

We also consider the effect of role hierarchies on the $\mathit{ESPOON_{ERBAC}}$ performance. In a role hierarchy, we assume that a role can inherit all permissions from its base role. This simplifies the role management and permission assignment to roles. In our experimentation, we consider 50 roles where each role has 5 permissions. Furthermore, there is a role hierarchy graph containing 25 roles, which is necessary for finding inheritance relationship between roles. Figure \ref{fig:espoon-vs-erbac} shows a very slight performance gain to evaluate the access request in case of role hierarchy in $\mathit{ESPOON_{ERBAC}}$. Since the permission can be associated with base role, we need to traverse in the role hierarchy graph to find base roles. The performance of traversing in the role hierarchy graph is shown in Figure \ref{fig:espoon-vs-erbac}. Finally, the requested permission is granted if associated even with any base roles. The role hierarchy may improve performance but in the worst case it incurs higher overhead. However, the performance of $\mathit{ESPOON_{ERBAC}}$ with role hierarchy is still better than that of $\mathit{ESPOON}$.

\section{Conclusions and Future Work}
\label{sec:conclusions-future-work}
In this paper, we have presented the $\mathit{ESPOON_{ERBAC}}$ architecture to support RBAC policies for outsourced environments. Our approach separates the security policies from the actual enforcing mechanism while guaranteeing the confidentiality of RBAC policies assuming the SP is honest-but-curious. The main advantage of our approach is that RBAC policies are encrypted but it still allows the PDP to perform the policy evaluation without revealing contents of requests or policies. Second, $\mathit{ESPOON_{ERBAC}}$ is capable of handling complex contextual conditions involving non-monotonic boolean expressions and range queries. Finally, the authorised users do not share any encryption keys making the process of key management very scalable. Even if a user key is deleted or revoked, the other entities are still able to perform their operations without requiring re-encryption of RBAC policies.

As future directions of our research, we are working on integrating a secure audit mechanism in $\mathit{ESPOON_{ERBAC}}$. The mechanism should allow the SP to generate genuine audit logs without allowing the SP to get information about both the data and the policies. However, an auditing authority must be able to retrieve information about who accessed the data and what policy was enforced for any access request made. Another direction of our work is towards the extension of the encrypted search and match capabilities to handle the case of negative authorisation policies and policies for long-lived sessions where the conditions need to be continuously monitored and the attributes of the request can be dynamically updated.

\section*{Acknowledgment}
The work of the first and third authors was supported by the EU FP7 research grant 257063 (project ENDORSE) while the work of the fourth author was supported by the Italian MIUR PRIN (project Autonomous Security).



%% References


% TODO: fix reference issues: title not visible in model1c-num-names, use other styles from Elservier reference templates.
%\section*{REFERENCES}
\bibliographystyle{plain} %model1c-num-names
\bibliography{cose-2012-espoon-erbac-references}


\clearpage

\section*{Vitae}

% left bottom right top
\parpic{\includegraphics[width=1.1in,height=1.5in,trim=135mm 110mm 10mm 40mm,clip]{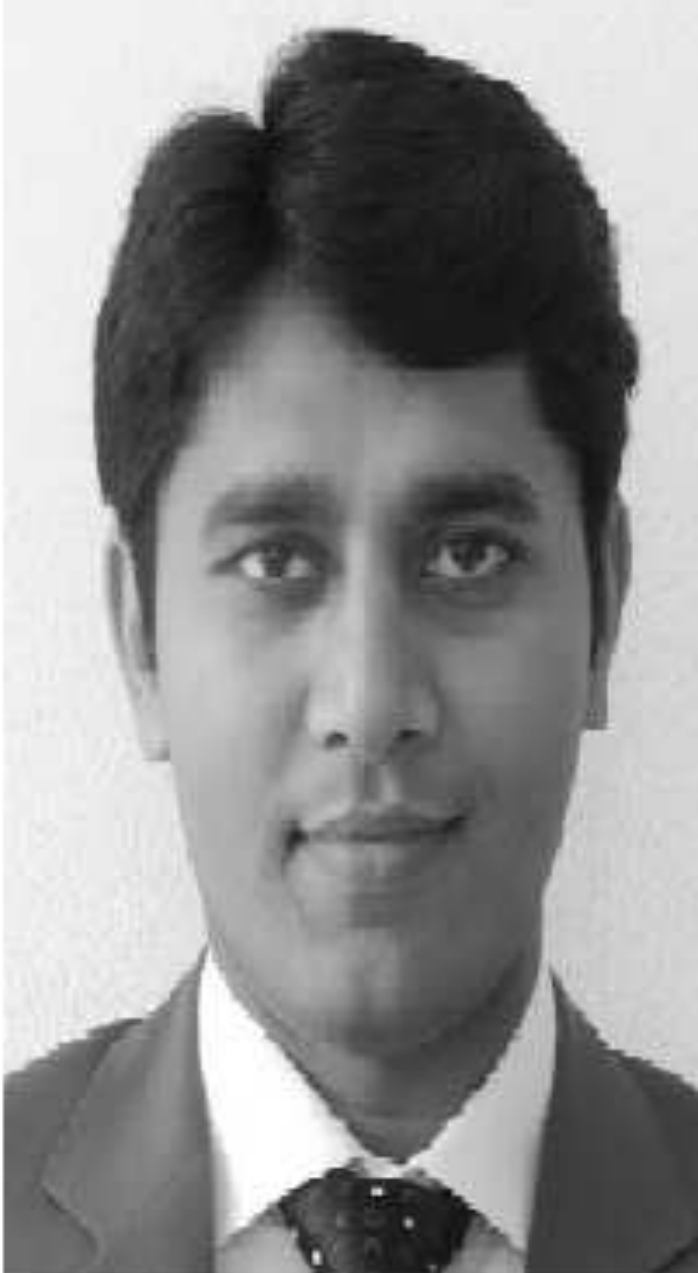}} % ,keepaspectratio
\noindent {\bf Muhammad Rizwan Asghar} received his B.Sc. (Hons.) degree in Computer Science from University of the Punjab, Lahore, Pakistan, in 2006. In 2009, he obtained his M.Sc. degree in Information Security Technology from Eindhoven University of Technology, the Netherlands. He joined Create-Net (an international research center based in Trento, Italy) in 2010. Currently, he is a Ph.D. candidate at University of Trento, Italy. His research interests include access controls, applied cryptography, cloud computing, security and privacy.

% left bottom right top
\parpic{\includegraphics[width=1.1in,height=1.5in,trim=145mm 120mm 0mm 20mm,clip]{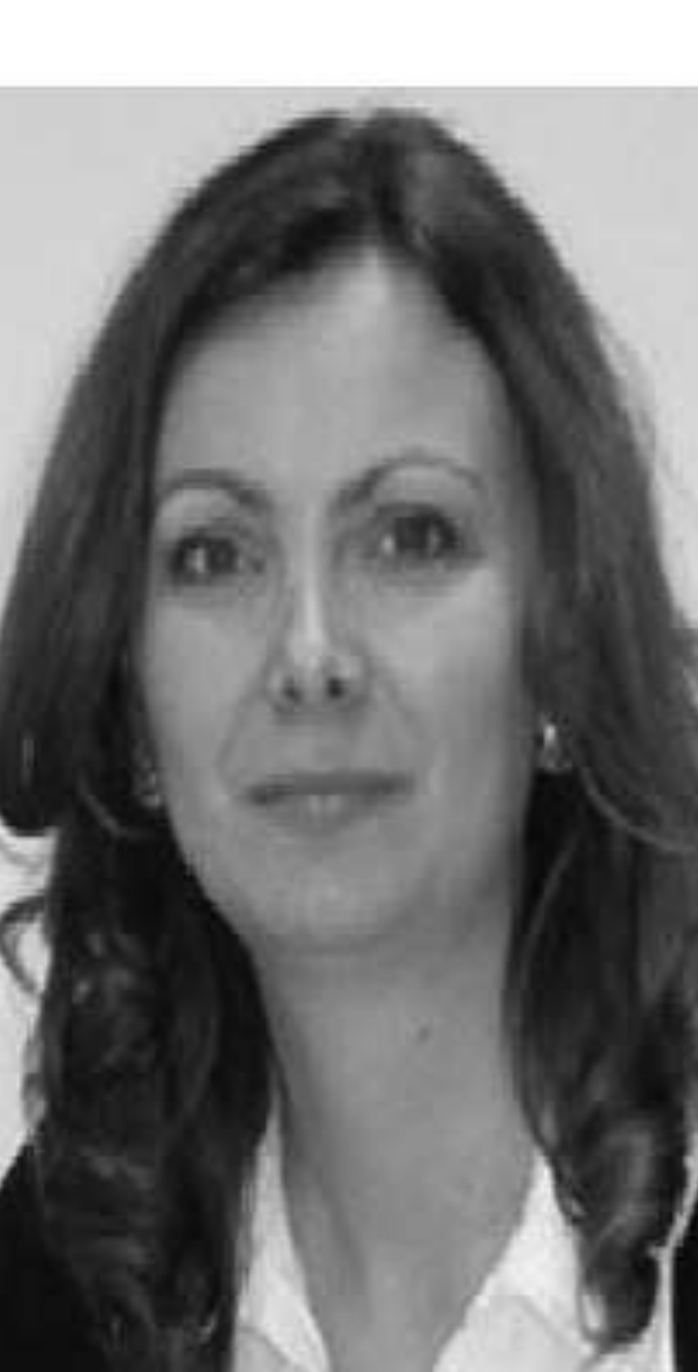}}
\noindent {\bf Mihaela Ion} received her B.Sc. in Information Technology and M.Sc. in Computer Science from International University in Germany. During her studies, she conducted various research projects with University of Marseille in France, SAP Waldorf and IBM Research Boeblingen in Germany. She joined CREATE-NET in 2007 where she's been working on various EU and Italian projects. Her research topics include data confidentiality in publish/subscribe systems, privacy for e-health applications, distributed identity and trust management. She is currently a Ph.D. candidate at the University of Trento working on security of publish/subscribe systems.

% left bottom right top
\parpic{\includegraphics[width=1.1in,height=1.5in,trim=145mm 120mm 0mm 20mm,clip]{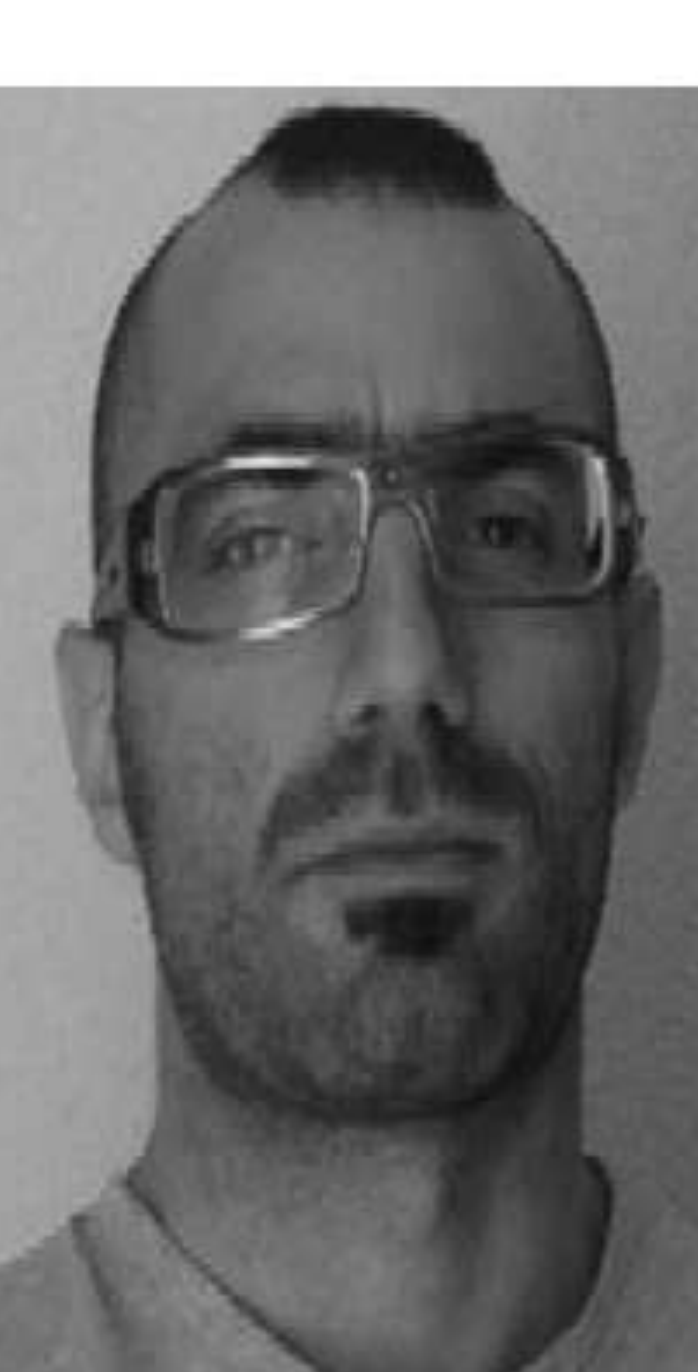}}
\noindent {\bf Giovanni Russello} is a lecturer at the University of Auckland, New Zealand, and leads the Security technical group within the iNSPIRE area at CREATE-NET in Trento, Italy. Giovanni received his M.Sc. (summa cum laude) in Computer Science from University of Catania, Italy in 2000. In 2006, he obtained his Ph.D. from the Eindhoven University of Technology. After obtaining his Ph.D., Giovanni moved to the Policy Group in the Department of Computing at Imperial College London. Giovanni's research interests include policy-based security systems, privacy and confidentiality in cloud computing, smartphone security, and applied cryptography.

% left bottom right top
\parpic{\includegraphics[width=1.1in,height=1.5in,trim=145mm 120mm 0mm 20mm,clip]{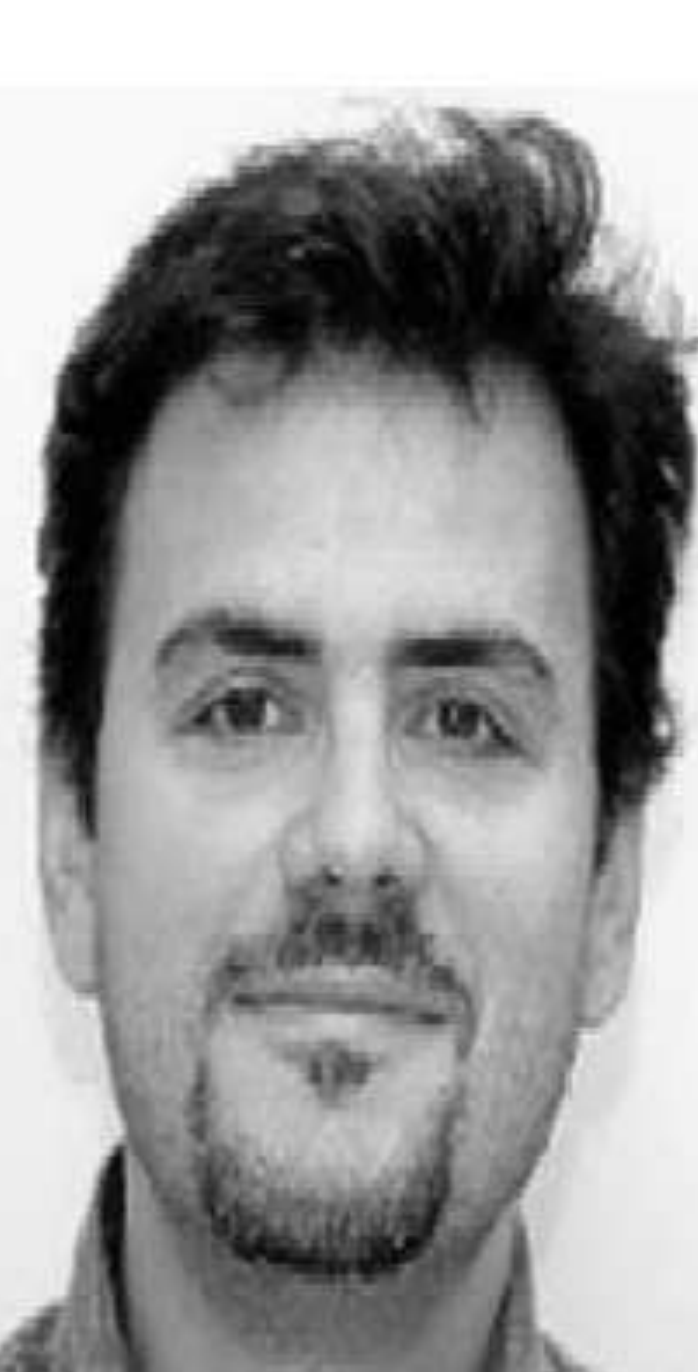}}
\noindent {\bf Bruno Crispo} received his Ph.D. in Computer Science from University of Cambridge, UK in 1999, having awarded a M.Sc. in Computer Science from University of Turin, Italy in 1993. He is an associate professor at University of Trento since September 2005. Prior to that he was Associate Professor at Vrije Universiteit in  Amsterdam, He is Co-Editor of the Security Protocol  International Workshop proceedings since 1997. He is member of ACM and senior member of IEEE. His main research interests spans across the field of security and privacy. In particular his recent work focus on the topic of security protocols, access control in very large distributed systems, distributed policy enforcement, embedded devices and smartphone security and privacy and privacy-breaching malware detection. He has published more than 100 papers in international journals and conferences on security related topics.

\end{document}